\documentclass[twocolumn]{aastex701}

\usepackage{amsmath}
\usepackage{booktabs}
\usepackage{threeparttable}
\usepackage{CJKutf8}
\defcitealias{Gaia2023}{Gaia Collaboration et al. 2023}
\begin{document}
\begin{CJK*}{UTF8}{gbsn}
\received{2026 May 14}
\revised{2026 June 5}
\accepted{2026 July 10}

%\title{Detectability of Oblateness and Exomoons in Exoplanetary Systems}
\title{How Many Transiting Giant Planets Can JWST Search for Moons and Rotational Oblateness?}

\author[0000-0002-6379-3816]{Le-Chris Wang~(王乐)}
\affiliation{Department of Astrophysical Sciences, Princeton University, 4 Ivy Lane, Princeton, NJ 08544, USA}
\email{lechris.wang@princeton.edu}

\author[0000-0002-4265-047X]{Joshua N.\ Winn~(温乔书)}
\affiliation{Department of Astrophysical Sciences, Princeton University, 4 Ivy Lane, Princeton, NJ 08544, USA}
\email{jnwinn@princeton.edu}

\correspondingauthor{Le-Chris Wang}
\email{lechris.wang@princeton.edu}

\begin{abstract}
\noindent Observations with the {\it James Webb Space Telescope} (JWST) can, in principle, detect
moons and rotational oblateness of giant exoplanets
through subtle distortions of transit light curves.
The most favorable planets are expected to be on wide orbits ($\gtrsim$0.3~AU)
where moons and rapid rotation are more likely to survive tidal evolution.
No unambiguous detections have yet been reported.
Here, we forecast the number of systems with sufficiently
favorable properties to allow for secure detections, using
JWST noise models, analytic detectability scalings,
giant-planet occurrence rates, and the Gaia star catalog.
For planets orbiting 0.9--1.6$\,M_\odot$ stars and
a noise model based on demonstrated JWST performance,
single-transit observations
should be capable of detecting Jupiter-like rotational oblateness
in several known systems and of order 10 systems yet to be discovered,
if obliquities are typically $\gtrsim$10$^\circ$.
A similar number of systems are favorable for
Ganymede-sized moons, if such moons are common.
The yields can increase to tens or hundreds of systems
if lower-mass host stars are included or if
JWST can achieve photon-limited performance.
Time-correlated noise on 1--10 hr timescales can strongly
suppress these yields; a noise floor of a few tens of parts per million
is enough to hide oblateness or moons in many otherwise favorable systems.
Successful searches
will therefore require both a more complete census of long-period transiting
giant planets and low levels of instrumental
systematics and stellar variability.
\end{abstract}


\section{Introduction}

At the highest precision, transit light curves
become sensitive to exoplanet properties
beyond the usual measurements of
planetary radii, orbital ephemerides,
and atmospheric spectra.
In principle, transit light curves can
reveal rotational oblateness \citep{seagerhui2002,barnesfortney2003},
planetary rings \citep{barnesfortney2004}, and moons
\citep{sartoretti1999}.
These phenomena produce subtle deviations
from the standard transit shape or, in the case of moons,
additional flux dips \citep[see Figure~2 of][]{millhollandwinn2025}.
For Jupiter-sized planets and moons with sizes
comparable to the Galilean satellites,
the expected signals are at the level of
10--100 parts per million (ppm).
Achieving this precision is difficult
but has become possible
with space telescopes, especially
the James Webb Space Telescope (JWST).

To date, however, there have been no unambiguous detections of 
oblateness, rings, or moons.
Upper limits on oblateness have been reported
by \cite{carterwinn2010,Zhu2014,Biersteker2017,lammerswinn2024,LiuZhu2024} and \cite{cassese2026}.
Searches for rings have been
carried out by \cite{heising2015,aizawa2018}, and \cite{umetani2025}.
Upper limits on moons have been obtained
by \cite{kipping2015, Kipping+2022, Kipping2025} and \cite{Pass2026},
although a few debated candidates remain
\citep{teachey2018,Kreidberg+2019, Kipping+2022}.

These null results do not necessarily imply
that future prospects for detecting these effects are dim.
They may instead reflect the small number of favorable
targets observed so far, or the fact that many
of the best systems that should exist 
have not yet been discovered.
The scientific motivation is strong
because these phenomena
probe planet properties that are
otherwise accessible only in the Solar System:
spin states and internal structure (rotational oblateness),
circumplanetary material (rings),
and satellite formation and evolution (moons).

The goal of this paper is to estimate how many transiting
planets are favorable targets for such searches
with JWST.
Will rotational oblateness and moons
remain out of reach except in rare special cases?
Or are there enough favorable systems, known or undiscovered,
to make population-level studies possible,
thereby establishing new links between exoplanet science
and
traditional planetary science?

This paper is organized as follows.
Section~\ref{sec:noise} assesses JWST photometric
performance, both in the photon-limited regime
and based on demonstrated performance in transit observations.
Section~\ref{sec:metrics}
presents simple
scaling relations for evaluating the detectability
of moons and oblateness; we do not separately consider
rings because the requirements are generally
similar to those for oblateness.
These scalings are applied to simulated populations
of transiting planets based
on a star catalog described in Section~\ref{sec:stellar_population_model}
and a planet-occurrence model described in 
Section~\ref{sec:planet_occurrence_model}.
Section~\ref{sec:detectable_systems} gives the
predicted yields of favorable
systems.
Section~\ref{sec:rednoise} examines the sensitivity
of those yields to the level of time-correlated noise
in light curves.
Section~\ref{sec:discussion}
compares these predictions with the current sample
of known transiting planets and discusses the main
limitations of the analysis.
Section~\ref{sec:conclusion} summarizes our conclusions.

\section{JWST Photometric Precision}\label{sec:noise}

\begin{figure}[t]
    \centering
    \includegraphics[width=1.\linewidth]{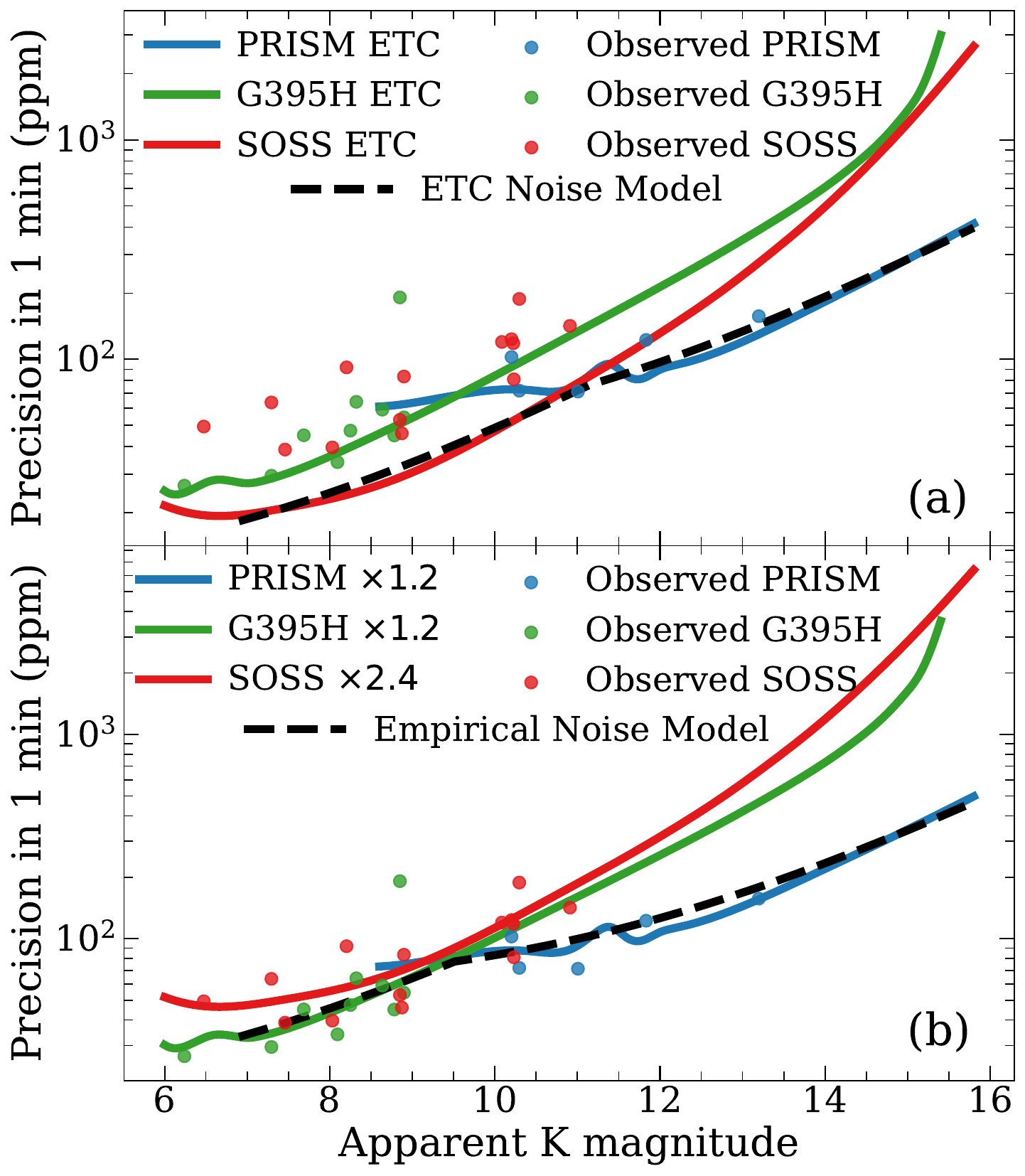}
    \caption{JWST photometric precision per minute,
    based on bandpass-summed
    time-series data and plotted versus apparent $K$ magnitude.
    \textit{(a)} Empirically measured scatter reported
    in the literature (points),
    compared with Exposure Time
    Calculator predictions from \texttt{PandExo}
    for three instruments/modes (curves).
    The dashed curve is a fitting function for
    the lower envelope of the ETC predictions
    (Equation \ref{eq:noise_etc}); in this
    fit, NIRISS/SOSS is favored for $K<11.2$ and
    NIRSpec/PRISM at fainter magnitudes.
    \textit{(b)} Same, but with curves 
    fitted to empirical scatter rather than ETC predictions
    (Equation~\ref{eq:noise_realistic});
    in this model, NIRSpec/G395H is favored for $K<9.5$
    and NIRSpec/PRISM at fainter magnitudes. }
    \label{fig:noise}
\end{figure}

We focus on JWST because it is
the best available facility for
ultraprecise photometry over long, continuous observing intervals.
For the bright targets most relevant to this study,
time-series spectroscopy
with NIRISS/SOSS and NIRSpec/BOTS
(using either PRISM or G395H) has demonstrated
the highest precision.
These modes have achieved bandpass-summed
(``white-light'') precisions of order
100~ppm per minute
in observations of planets such as WASP-39\,b \citep{feinstein2023,rustamkulov2023,Alderson2023}.
Below, we evaluate the relevant photometric precision in two ways:
first from the Exposure Time Calculator (ETC) predictions,
and then from empirically measured residual scatter
in published transit observations.
We do not attempt a comparison with the other two
JWST instruments, MIRI and NIRCam, which are generally
less competitive for this application due to lower photon
rates or throughput, except possibly in special cases\footnote{MIRI, for example, has the advantage of suppressed stellar activity signals in its bandpass.}.

\subsection{Exposure Time Calculator Predictions}

To estimate the best-case noise level
for JWST time-series observations,
we used \texttt{PandExo}\footnote{\href{https://github.com/natashabatalha/PandExo}{https://github.com/natashabatalha/PandExo}.} \citep{batalha2017},
which interfaces with \texttt{Pandeia} \citep{pontoppidan2016}.
For apparent $K$ magnitudes from 6 to 16,
we used \texttt{PandExo} to select the most efficient non-saturating
observing configuration and to return the corresponding
cadence and expected signal-to-noise ratio.
We evaluated three modes, NIRISS/SOSS,
NIRSpec/PRISM, and NIRSpec/G395H.
For simplicity, we adopted a representative
stellar spectrum with $T_{\rm eff}=5100~\mathrm{K}$ and $\log g = 4.5$ typical of a 0.9~$M_\odot$ star, near the middle of the range of stars we considered.

The resulting photometric precision, expressed as the expected scatter in 1-minute bins,
is shown in Figure \ref{fig:noise}(a).
NIRISS/SOSS provides the best
predicted precision for the brightest targets
($K\lesssim 11.2$), while
NIRSpec PRISM is best at fainter magnitudes.
For use in the analytic forecasts below,
we approximate the lower envelope of the ETC predictions with
the following piecewise function:
\begin{equation} \frac{\sigma_{\rm ETC}}{\rm ppm} =
\begin{cases}
70\cdot 10^{0.2\,(K-11.2)} + 9, & 7 \leq K \le 11.2, \\
44\cdot 10^{0.2\,(K-11.2)} + 34, & 11.2 < K < 16.
\end{cases}
\label{eq:noise_etc}
\end{equation}
These formulas give the estimated precision
when employing the optimal instrument mode among those considered.
Extrapolation to $K<7$ is probably
invalid because such bright targets
would saturate the NIRISS detector in the configurations considered here.

\subsection{Empirical Noise Model}

\begin{table*}[t]
\centering
\caption{White-light photometric scatter per 1-min bin,
as predicted by the ETC and as measured.}
\label{tab:wl_scatter_vs_etc}
\small
\setlength{\tabcolsep}{6pt}
\renewcommand{\arraystretch}{1.0} % {1.15}
\begin{tabular}{llcccl}
\toprule
Planet & Instrument & $\sigma_{\rm ETC}$ & $\sigma_{\rm obs}$ & $\sigma_{\rm obs}/\sigma_{\rm ETC}$ & Reference \\
 & &  (ppm) & (ppm) & &\\
\midrule
HAT-P-18 b      & NIRISS/SOSS     & 54  & 81  & 1.5 & \cite{FT2024} \\
WASP-69 b       & NIRISS/SOSS     & 24  & 39  & 1.6 & L.-C.\ Wang et al., in preparation \\
WASP-94 A b     & NIRISS/SOSS     & 28  & 46  & 1.7 & \cite{mukherjee2025} \\
WASP-166 b      & NIRISS/SOSS     & 22  & 40  & 1.8 & \cite{mayo2025} \\
HAT-P-14 b      & NIRISS/SOSS     & 27  & 53  & 2.0 & \cite{LiuWang2025} \\
WASP-96 b       & NIRISS/SOSS     & 71  & 142 & 2.0 & \cite{Wang2025} \\
WASP-52 b       & NIRISS/SOSS     & 50  & 120 & 2.4 & \cite{FT2025} \\
WASP-39 b       & NIRISS/SOSS     & 51  & 124 & 2.4 & \cite{feinstein2023} \\
WASP-17 b       & NIRISS/SOSS     & 48  & 118 & 2.5 & \cite{louie2025} \\
K2-18 b         & NIRISS/SOSS     & 33  & 84  & 2.5 & \cite{schmidt2025} \\
GJ-357 b        & NIRISS/SOSS     & 17  & 49  & 2.9 & \cite{taylor2025} \\
TRAPPIST-1 c    & NIRISS/SOSS     & 64  & 188 & 3.0 & \cite{radica2025} \\
GJ-3090 b       & NIRISS/SOSS     & 20  & 64  & 3.2 & \cite{ahrer2025} \\
TOI-732 c       & NIRISS/SOSS     & 25  & 92  & 3.7 & \cite{rigby2025} \\
\hline
TOI-5205 b      & NIRSpec/PRISM   & 95  & 71  & 0.8 & \cite{canas2025} \\
TRAPPIST-1 e    & NIRSpec/PRISM   & 71  & 72  & 1.0 & \cite{espinoza2025} \\
Kepler-51 d     & NIRSpec/PRISM   & 126 & 157 & 1.2 & \cite{lammerswinn2024} \\
WASP-39 b       & NIRSpec/PRISM   & 76  & 102 & 1.4 & \cite{rustamkulov2023} \\
Kepler-167 e    & NIRSpec/PRISM   & 86  & 123 & 1.4 & \cite{Kipping2025} \\
\hline
V1298 Tau b     & NIRSpec/G395H   & 38  & 34  & 0.9 & \cite{barat2025} \\
GJ-1214 b       & NIRSpec/G395H   & 44  & 45  & 1.0 & \cite{schlawin2024} \\
GJ-3090 b       & NIRSpec/G395H   & 28  & 30  & 1.1 & \cite{ahrer2025} \\
K2-18 b         & NIRSpec/G395H   & 48  & 54  & 1.1 & \cite{schmidt2025} \\
GJ-486 b        & NIRSpec/G395H   & 22  & 27  & 1.2 & \cite{moran2023} \\
WASP-107 b      & NIRSpec/G395H   & 46  & 59  & 1.3 & \cite{sing2024} \\
TOI-270 d       & NIRSpec/G395H   & 37  & 47  & 1.3 & \cite{holmberg2024} \\
LHS-475 b       & NIRSpec/G395H   & 29  & 45  & 1.5 & \cite{LY2023} \\
GJ-1132 b       & NIRSpec/G395H   & 37  & 64  & 1.7 & \cite{Bennett2025} \\
HAT-P-14 b      & NIRSpec/G395H   & 51  & 191 & 3.7 & \cite{Espinoza2023} \\
\bottomrule
\end{tabular}
\normalsize
\end{table*}

As a reality check on the ETC predictions, we compiled measurements
from the literature of white-light residual scatter in JWST
transit observations.
Table~\ref{tab:wl_scatter_vs_etc} gives
the measured scatter in 1-minute bins, the corresponding
ETC prediction, and their
ratio. These measurements are heterogeneous; they differ
in stellar properties, observing setup, wavelength
range, and reduction method, 
among other factors. Thus, the compilation should be interpreted
as providing a practical benchmark rather than a highly accurate
instrument model.

As shown in Figure \ref{fig:noise}(a),
the empirical precision is typically worse than
the ETC prediction
by median factors of 1.2, 1.2, and 2.4 for
NIRSpec/PRISM, NIRSpec/G395H, and NIRISS/SOSS,
respectively. 
The larger discrepancy for
NIRISS/SOSS may reflect imperfect correction
of $1/f$ noise, because the bright spectral trace
leaves relatively few unilluminated pixels
for background estimation \citep{radica2023, Albert2023}.
Based on the current literature,
NIRISS/SOSS generally appears to be
less precise than NIRSpec/G395H for bright targets.

We therefore constructed an empirical noise model using
NIRSpec/G395H for bright targets and NIRSpec/PRISM for fainter targets:
\begin{equation}
\frac{\sigma_{\rm emp}}{{\rm ppm}} =
\begin{cases}
63\cdot 10^{0.2(K-9.5)} + 14, & 7 \leq K \le 9.5, \\
23\cdot 10^{0.2(K-9.5)} + 55, & 9.5 < K < 16.
\end{cases}
\label{eq:noise_realistic}
\end{equation}
The empirical model is plotted in Figure \ref{fig:noise}(b). It reproduces the trend in the measurements drawn from the literature with typical fractional differences of about $20\%$.
This model should be understood as a description
of actual performance based on current practices and data analysis methods,
not as a fundamental limit.
Improvements in calibration, detrending, and instrument
characterization might lead to significant improvements and
a closer approach to the ETC predictions.

Although the JWST noise models are expressed
in terms of 1-minute bins, the relevant signals have characteristically longer timescales. For oblateness, the relevant timescale is typically of order 0.5~hr, about half of the ingress/egress duration. For moon detection, the relevant timescale is typically from several to ten hours, about half of the full transit duration. The 1-minute precision is therefore only a convenient normalization for a white-noise model, in which the fluctuation level is assumed to decrease with the square root of the averaging time. This model might not fully capture the noise properties on the actual signal timescales, an issue discussed in Section \ref{sec:rednoise}. 

\section{Simple Detectability Metrics}
\label{sec:metrics}

We now derive simple metrics for estimating whether rotational
oblateness or a moon would leave
a detectable imprint in transit photometry.
The goal is not to replace a full light-curve analysis,
but to obtain scaling relations that can be evaluated
for large simulated populations using
quantities that are generally known in advance:
stellar properties, orbital geometry, and photometric
precision. We quantify detectability by the expected
improvement in $\chi^2$ when
a model including oblateness or a moon is compared with the
best-fitting standard transit model. Assuming independent Gaussian uncertainties, $\sqrt{\Delta\chi^2}$ is the model-comparison signal-to-noise ratio \citep{Kipping2023}.

\subsection{Oblateness}\label{subsec:oblateness_metric}

\begin{figure*}[t]
    \centering
    \includegraphics[width=1.0\linewidth]{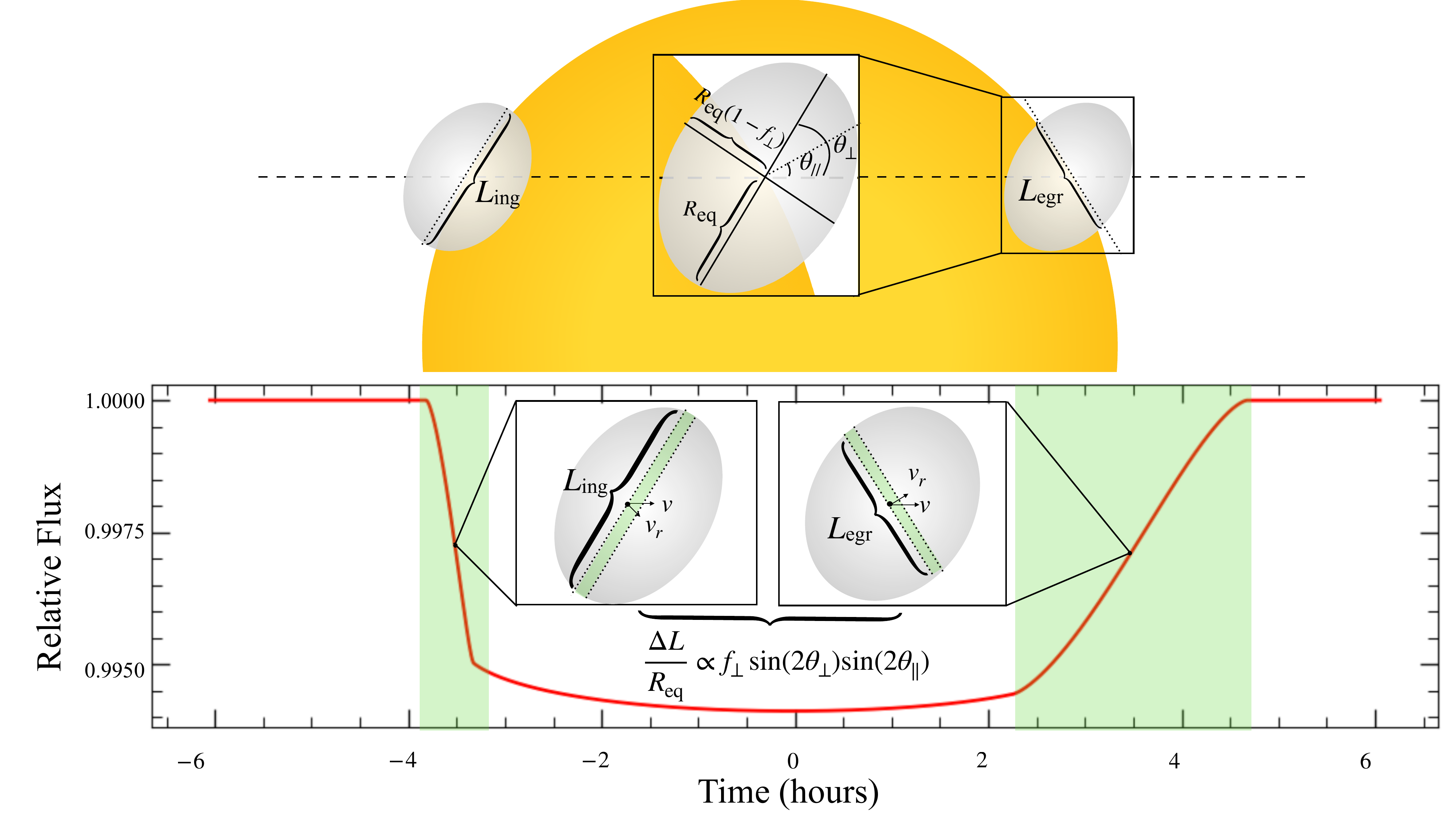}
    \caption{Illustration of the ingress--egress
    asymmetry produced by an oblate planet.
    \textit{Top:} Transit geometry. The planet's projected shape
    is elliptical, with projected flattening $f_\perp$. The intersections of the ellipse with the
    stellar limb are nearly straight lines but with
    different lengths $L$, leading to different
    rates of ingress and egress.
    \textit{Bottom:} Example light curve (red) with
    exaggerated projected oblateness ($f_\perp = 0.8$).
    The resulting asymmetry between ingress and egress (green)
    underlies the geometric factor $F$ used in our detectability
    metric (Appendix~\ref{app:G-factor-derivation}).
    }
    \label{fig:geometry}
\end{figure*}

A planet's rotational flattening is set mainly by the ratio of
centrifugal to gravitational acceleration, and therefore increases
as the square of the rotation rate.
Jupiter and Saturn are
spinning quickly, with rotation periods
of $\approx$10~hours, and their polar radii
are shorter than their equatorial radii
by 6.5\% and 10\%, respectively.
However, most of the known transiting giant
planets are on close orbits for which tidal
interactions are expected to have driven
them into spin-orbit synchronization, reducing
rotational flattening to
a negligible level \citep{carterwinn2010}.
We therefore restrict our analysis to giant planets
with periastron distances larger than 0.3~AU, for
which tidal despinning is expected to
be inefficient and rapid rotation can plausibly persist
\citep{carter2010}.\footnote{The threshold of 0.3~AU
was obtained 
by requiring the tidal spindown timescale
$\tau_{\rm s} \gtrsim 13.8$~Gyr
using Equation~12 of \cite{carter2010},
assuming Jupiter's mass, radius, and rotation period,
a moment-of-inertia constant $C=0.25$, and a
tidal quality factor $Q'_p=10^{5.5}$.
See also Section~\ref{sec:discussion}.}

We describe the sky-projected shape of an oblate planet
by its projected flattening $f_\perp$ and projected
obliquity $\theta_\perp$. Here, $1-f_\perp$
is the axis ratio of the projected ellipse,
and $\theta_\perp$ is the position angle of the long axis
of the projected ellipse relative to the transit chord.
Oblateness mainly affects the
ingress and egress phases of the light curve
and produces a characteristic asymmetry
as long as $\theta_\perp$ is not near $0^\circ$,
$90^\circ$, or $180^\circ$ (Figure~\ref{fig:geometry}).
This ingress-egress
asymmetry is the most diagnostic signature of oblateness
because it cannot be mimicked by small changes
in the usual transit parameters.

%The detectability of this signal depends on the degree of asymmetry, the duration $\tau$ of ingress and egress, the transit depth $\delta$, and the photometric precision $\sigma$. 
The residuals between an oblate-planet light curve and the best-fitting spherical model have a characteristic amplitude of order $\delta F$, where $\delta$ is the transit depth and $F$ is the fractional ingress--egress asymmetry. The signal is concentrated during ingress and egress, with duration $\tau$. For independent measurements with precision $\sigma$, $\Delta\chi^2$ scales as the sum of squared residuals:
% On general grounds, we expect
\begin{align}
\label{eq:Delta_BIC_Oblateness_1}
    \Delta \chi^2 \propto 
    \frac{\delta^2 \tau}{\sigma^2}\, 
    F(f_\perp, 
    \theta_{\perp}, b)^2.
\end{align}
Appendix~\ref{app:G-factor-derivation} shows
that, to leading order in $f_\perp$,
\begin{align}
    F(f_\perp,\theta_\perp,b) \propto f_\perp\,\sin 2\theta_\perp\,\sin 2\theta_\parallel,
\end{align}
where $\theta_\parallel = \sin^{-1} b$
is the angle between
the transit trajectory
and the local normal to the
stellar limb
(Figure~\ref{fig:geometry}).
Thus, as expected, $F$ vanishes
when $\theta_\perp$
is $0^\circ$,
$90^\circ$, or $180^\circ$.
Note, too, that 
$F$ is maximized
when $\theta_\parallel$ is
45$^\circ$ and $b=1/\sqrt{2}$.
This explains the earlier
finding based on detailed
simulations that
the optimal impact parameter
is near 0.7 \citep{barnesfortney2003, Zhu2014}.

Using standard small-planet transit scalings for
a circular orbit,
Equation \ref{eq:Delta_BIC_Oblateness_1}
can be cast in terms of
the orbital distance $a$,
planet radius $R_p$,
stellar mass $M_\star$,
and stellar radius $R_\star$:
\begin{align}
    \Delta \chi^2 &\propto
    \frac{R_p^5}{R_\star^4}
    \sqrt{\frac{a}
    {M_\star(1-b^2)}}
 \frac{F(f_\perp,
 \theta_{\perp}, b)^2}{\sigma^2}. \label{eq:oblateness_general_circular}
\end{align}
This scaling highlights the most important trends:
detectability increases strongly with
planet radius,
is enhanced for small host stars,
and benefits from wider orbits
through longer ingress and egress durations.
The factor $(1-b^2)^{-1/2}$ reflects the
lengthening of ingress and
egress as the impact parameter increases, although
the approximation eventually breaks down
for grazing transits.

\subsection{Moons}\label{subsec:moon_metric}

Although it is reasonable to expect most wide-orbiting giant planets 
to rotate as fast as Jupiter and Saturn,
the existence of large moons
cannot be taken for granted.
Thus, our moon detectability metric
specifies whether a moon of a specified radius
would be detectable if present, and our forecast
should be interpreted as a number of ``moon-searchable systems.''
An actual prediction of moon detections would
require assumptions about the occurrence rate of moons
as a function of radius and period.

Long-lived moons can only exist between
the Roche limit and a fraction of the planet's Hill sphere,
with the allowed region further modified by tidal
evolution \citep{Domingos2006,rosariofranco2020, dobos2021,kisare2024}. For close-in planets this region is small, and large moons are unlikely
to survive on Gyr timescales. 
Rapidly rotating planets may also provide
more favorable tidal conditions for
retaining moons \citep{tokadjian_piro2020}.
On an empirical level, the large regular moons of the Solar System
mostly have orbital distances smaller than 0.05 Hill radii. % 39/42 satellites
For a solar-mass star, the Roche limit equals 0.05 Hill radii
for a planetary orbital distance of 0.3~AU.
Motivated by these considerations,
we restrict the moon analysis to planets
with periastron distances larger than 0.3~AU,
the same criterion as for oblateness.

The simplest
photometric signature of a transiting moon is a
shallow transit dip
offset in time from the planetary
transit.\footnote{Other possible signals arising
from the planet's reflex motion, such as transit-timing
and transit-duration variations, are not considered here.}
An accurate moon model would have many parameters, and
overlapping transits of planets and moons
can produce complex light-curve shapes.
For a simple detectability metric,
we consider the favorable case in which the moon
transit is wholly separated in time
from the planet transit.
A more realistic approach would
account for the loss
of information due to overlap between the planet and moon
transits; for example, \cite{Kipping2021} described a
more conservative approach in which the effective signal-to-noise ratio is based only on the non-overlapping segment of the
moon transit. We chose our simpler metric to avoid making additional assumptions about satellite orbital distances,
periods, and phases. Thus, our moon detectability
metric should be interpreted as optimistic.

When the moon signal
is an independent transit
of depth $\delta_m = (R_m/R_\star)^2$ and duration $T$,
we expect
\begin{align}
\Delta \chi^2 \propto 
\frac{\delta_m^2 T}{\sigma^2} \propto
\left(\frac{R_m}{R_\star}\right)^{\!\!4}
\frac{T}{\sigma^2},
\label{eq:bic_moon_1}
\end{align}
For simplicity, we assume $T$ is approximately
the same as the planet's transit duration.
For a circular planetary orbit, this gives
\begin{align}
\Delta \chi^2 \propto 
\frac{R_m^4}{R_\star^3\,\sigma^2}
\sqrt{\frac{a(1-b^2)}{M_\star}}.
\label{eq:bic_moon_2}
\end{align}

The scaling relations for $\Delta\chi^2$
were validated with injection–recovery simulations,
which were also used to set the constants
of proportionality used in the yield calculations.
Details are provided in
Appendix~\ref{app:detectability_scalings}.

\section{Stellar Population Model}
\label{sec:stellar_population_model}

\begin{figure*}[t!]
    \centering
    \includegraphics[width=1.0\linewidth]{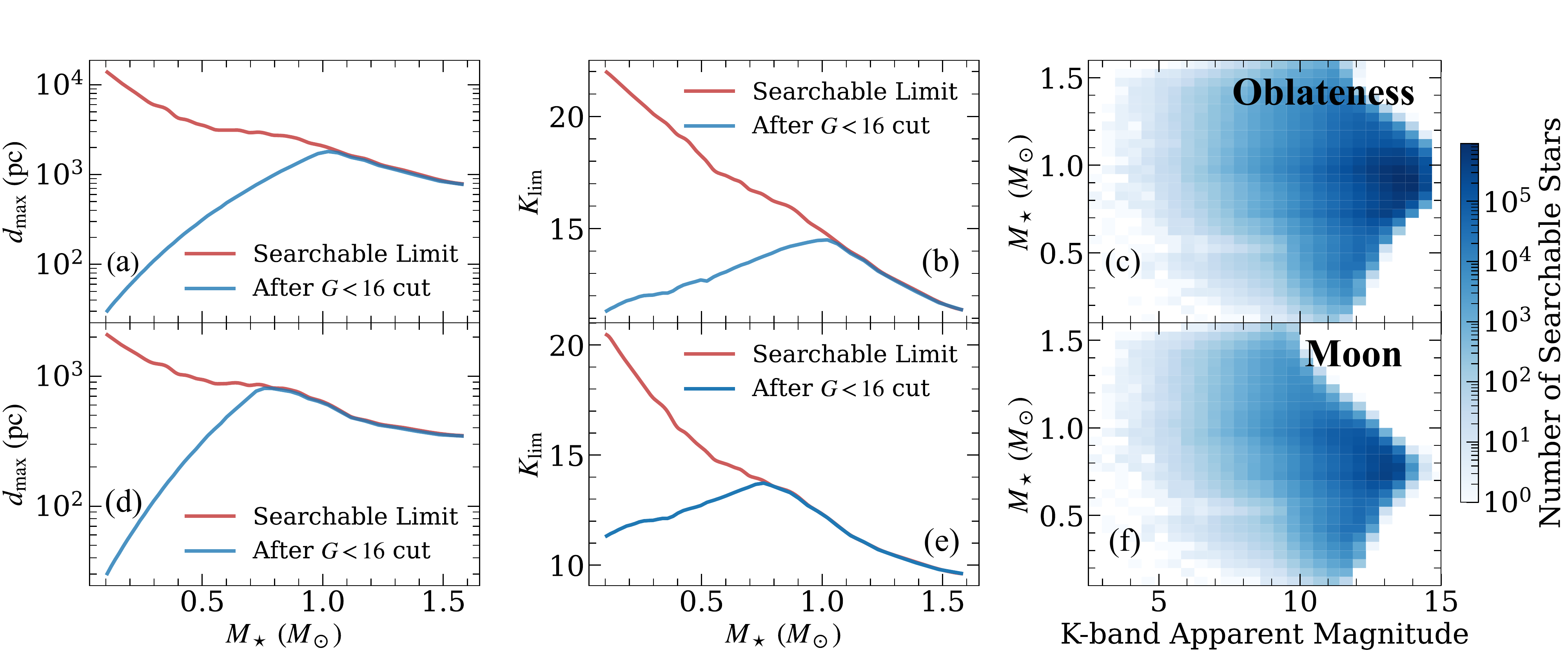}
    \caption{Favorable-star catalogs for detecting planetary oblateness (top) and moons (bottom) constructed under optimistic assumptions about planet/moon properties, viewing geometry,
    and JWST performance. The adopted detection threshold
    was $\Delta\chi^2=60$ using
    the ETC noise model.
    Panels a and d show the maximum search distance as a function of stellar mass.
    The solid curves show the formal
    distance limits, and the dashed
    curves show the revised limits
    after applying the $G<16$ cut.
    Panels b and e show the corresponding limiting apparent $K$ magnitudes. Panels c and f
    show the number of favorable stars in the $(M_\star, K)$ plane.
    %Integrating over the catalog yields $\sim12.1$ ($\sim1.5$) million favorable stars for oblateness (moons) for $M_\star>0.9~M_\odot$, and $\sim8.7$ ($\sim5.2$) million for $M_\star<0.9~M_\odot$.
    }
    \label{fig:star_cat_oblateness}
\end{figure*}

Before assigning simulated planets to stars,
we first constructed catalogs of stars
for which the relevant signals might plausibly be detectable
if a favorable transiting planet were present.
For each star, we asked whether a fiducial Jupiter-like
planet or Ganymede-sized moon would produce a detectable
signal ($\Delta\chi^2$~$>60$, see Section \ref{sec:planet_occurrence_model}), given the star's
radius, mass, and apparent magnitude.
To ensure that the catalogs include
all potentially favorable host stars, we evaluated the
detectability metrics under optimistic
assumptions: for oblateness, 
$b = 1/\sqrt{2}$ and $\theta_{\perp} = 45^\circ$;
and for moons, a long transit duration corresponding
to $a=20$~AU.
We also used the ETC-based noise model rather than
the empirical noise model.
These inclusive star catalogs were the starting points
for Monte Carlo simulations that account for
less favorable geometries, planet occurrence rates,
and transit probabilities.

For each stellar mass, we computed the maximum searchable distance
$d_{\rm max}(M_\star)$ from the condition $\Delta\chi^2 = 60$,
using the scaling relations from Section~\ref{sec:metrics}
with $K=M_K + 5\log(d/10\,{\rm pc})$.
The resulting limits show that detectability improves strongly toward
lower-mass, smaller stars, which can be searched to larger distances despite
their lower luminosities (Figure~\ref{fig:star_cat_oblateness}).
At the lowest stellar masses, however, the formal search volume extends
to several kpc, where extinction, crowding, and
transit-survey incompleteness become important.
Giant-planet occurrence is also lower and more uncertain
than for Sun-like stars.
We therefore considered two stellar samples:
a baseline sample with $0.9 < M_\star/M_\odot < 1.6$,
and a lower-mass sample spanning $0.1< M_\star/M_\odot< 0.9$.
For the lower-mass sample we imposed an
additional cut of $G<16$, intended to exclude faint systems
unlikely to be discovered by current or near-future
transit surveys. The effect of this magnitude limit
on the search distance is shown in Figure \ref{fig:star_cat_oblateness}.

We constructed the star catalogs using data from Gaia DR3 \citepalias{Gaia2023},
selecting stars within $d<d_{\rm max}$ and restricting the sample
to likely main-sequence stars.
After correcting for extinction, we required
$0.4 < M_G - 2.9(G_{BP} - G_{RP}) < 4.5$,
following \cite{lammerswinn2026}.
We used the Gaia-provided extinction corrections when available,
and otherwise estimated the correction from tabulated results
for nearest-neighboring stars
in Galactic longitude, latitude, and distance. We also
required \texttt{RUWE} $<1.2$ to reduce
contamination from unresolved binaries; this cut removed
about 20\% of the stars.

The resulting oblateness-hunting star catalog
contains about 12.1 million stars with
masses between 0.9 and 1.6~$M_\odot$ and
8.7 million stars with masses between 0.1 and 0.9~$M_\odot$.
For the moon-hunting catalog, the
corresponding totals are
1.5 million and 5.2 million stars.

Stellar masses and radii were taken from
the Gaia DR3 astrophysical-parameters
table \citep{creevey2023} whenever possible
(about 60\% of the sample). For
the remaining stars, we estimated masses and radii
from zero-age main-sequence (ZAMS) relations
based on the extinction-corrected $M_G$ \citep{PecautMamajek2013}.
For stars with Gaia-reported masses and radii,
these relations have median absolute deviations
of 0.07~$M_\odot$ and 0.04~$R_\odot$.
We obtained $K$-band magnitudes
from 2MASS when possible (about 20\% of the sample),
and otherwise estimated them
from the ZAMS $G-K$
relation, neglecting $K$-band extinction.
The rightmost
column of Figure \ref{fig:star_cat_oblateness}
shows the resulting
distribution of searchable
stars as a function of $M_\star$ and $K$.

\section{Planet Occurrence Model}
\label{sec:planet_occurrence_model}

\begin{figure*}[t!]
    \centering
    \includegraphics[width=1.0\linewidth]{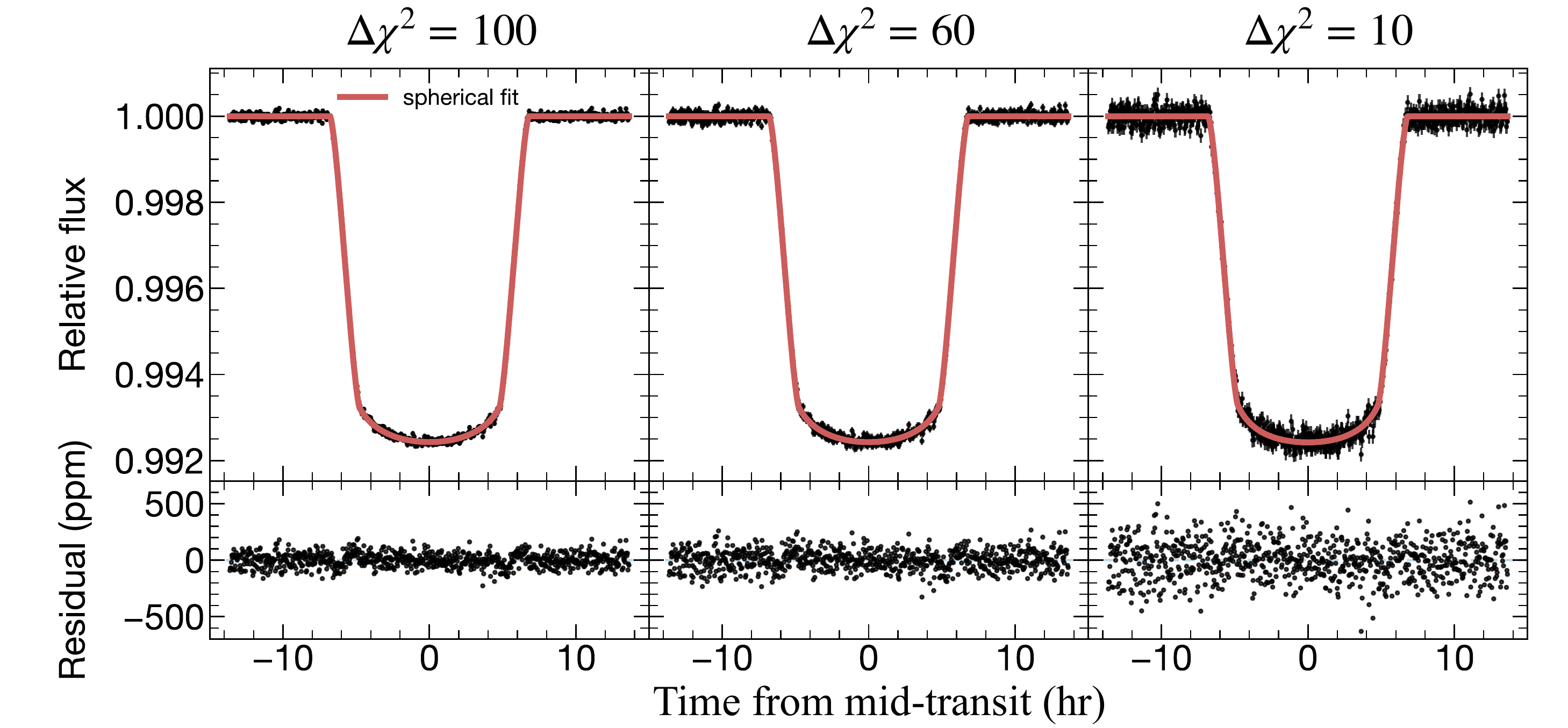}
    \caption{{ Simulated light curves illustrating
    different levels of detection significance.
    The three columns show cases with $\Delta\chi^2=100$,
    60, and 10 between the oblate model and the best-fit standard
    model with zero oblateness.
    \textit{Top row:} Noisy transit photometry (black) and the best-fitting standard
    model (red).
    \textit{Bottom row:} Residuals.
    We adopted $\Delta\chi^2\gtrsim 60$ as the fiducial
    detectability threshold, and show the dependence
    of the yields on this threshold in Figure~\ref{fig:properties_oblateness}.}}
    \label{fig:delta_bic}
\end{figure*}

We next populated the favorable-star catalogs
with hypothetical giant planets.
We adopted the giant-planet occurrence model from the California Legacy Survey (CLS) \citep{rosenthal2021,Fulton2021}. The occurrence rate density is
\begin{equation}
\frac{dN_p}{dN_\star\, d\ln a\, d\ln m_p}=C\left(\frac{a}{\mathrm{AU}}\right)^{\!\beta}
\left[1-e^{-(a/a_0)^{\gamma}}\right],
\label{eq:occurrence_cls}
\end{equation}
for planet masses between $30$ and $6000~M_\oplus$ and semi-major axes $a$ between $0.1$ and $30$~AU. In each Monte Carlo realization,
we drew the parameters $(C, \beta, a_0, \gamma)$ from the CLS posterior
and assigned planets to stars assuming a Poisson process in $(\ln a, \ln m_p)$. When more than one planet
was assigned to a star, we retained only one.

For the lower stellar mass sample (0.1--0.9\,$M_\odot$),
we allowed the occurrence-rate
normalization to depend on stellar mass,
following \cite{lammerswinn2026}:
\begin{equation}
\tilde{C}(M_\star)=
C\left(\frac{M_\star}{k\,M_\odot}\right)^{\eta}.
\label{eq:occurrence_mass_general}
\end{equation}
We drew $\eta$ from $\mathcal{U}(1,2)$
and $k$ from $\mathcal{U}(0.8,1.0)$.\footnote{\cite{lammerswinn2026}
drew $\eta$ from $\mathcal{U}(0,2)$;
our choice is meant to be conservative,
to avoid overestimating the
occurrence rate of giant planets around lower-mass stars.
For $1< m/M_J < 13$, $0.1<M_\ast<0.6$,
and $0.1<a/{\rm AU}<20$,
this prescription
gives an occurrence rate
of $5.8^{+1.9}_{-1.4}\%$, consistent
with the $6.5\pm3.0\%$ rate reported by \cite{montet2014}.}

%$a$ from Eqn.~\ref{eq:occurrence_cls},

%We then restricted $a$ to $[0.3, 20]$~au, with the lower limit chosen to avoid tidal
% despinning and the (somewhat arbitrary) upper limit chosen because of the low transit probability
% at wide separations.

For each assigned planet, we drew $m_p$ from a log-uniform distribution
between 0.3 and 13\,$M_J$,
eccentricity from ${\rm Beta}(0.85, 2.27)$ \citep{Rosenthal2024},
and argument of pericenter
from $\mathcal{U}(0,2\pi)$. We drew $a$ from Equation~\ref{eq:occurrence_cls}, restricting $a(1-e)$ from 0.3 to 20 AU with the lower limit chosen to avoid tidal despinning and the (somewhat arbitrary) upper limit chosen because of the low transit probability
at wide separations.
We drew $\cos i$ from $\mathcal{U}(-1,1)$ and applied the geometric transit criterion, including the dependence on $e$ and $\omega$. For transiting systems, we evaluated the
detectability metrics
from Section~\ref{sec:metrics},
excluding grazing transits
for which the scaling relations
are inaccurate.

For the moon calculation, we assumed
a planet radius of 1~$R_J$ and
a Ganymede-sized moon of radius 0.037~$R_J$.
For the oblateness calculation, we
assumed a planet radius of 1~$R_J$
and an intrinsic flattening of
0.065.\footnote{We considered allowing
oblateness to depend
on planet mass. However, to leading order,
$f\propto \Omega^2 R_p^3/(G M_p)$ \citep[e.g.,][]{murraydermott1999},
and the empirical mass--rotation trend 
is approximately $\Omega \propto M_p^{1/2}$, suggesting that $f$ is nearly independent of $M_p$
at fixed $R_p$ \citep[e.g.,][]{Hughes2003,Scholz2018}.}
The azimuthal angle of the spin
axis around the orbit normal was drawn at
random from $\mathcal{U}(0,2\pi)$.

The distribution of obliquities
is unknown and is probably the largest
source of uncertainty in the forecasted yield of oblateness detections.
We therefore considered three choices:
(1) an isotropic distribution, with 
$\cos(\theta)$ drawn
from $\mathcal{U}(0,1)$;
(2) a Rayleigh distribution peaking
at $10^{\circ}$; and
(3) a Rayleigh distribution peaking
at Jupiter's obliquity of $3^{\circ}$.
We classified a system as yielding a
detection when $\Delta\chi^2>60$. 
Figure~\ref{fig:delta_bic} illustrates
representative oblate-planet light curves
with $\Delta\chi^2 = 10$, 60, and 100.

% We repeated this
% procedure for $10^4$ Monte Carlo realizations to obtain distributions of
% detection yields and system properties.

\section{Yield of Detections}
\label{sec:detectable_systems}

\begin{figure*}[t!]
    \centering
    \includegraphics[width=1.0\linewidth]{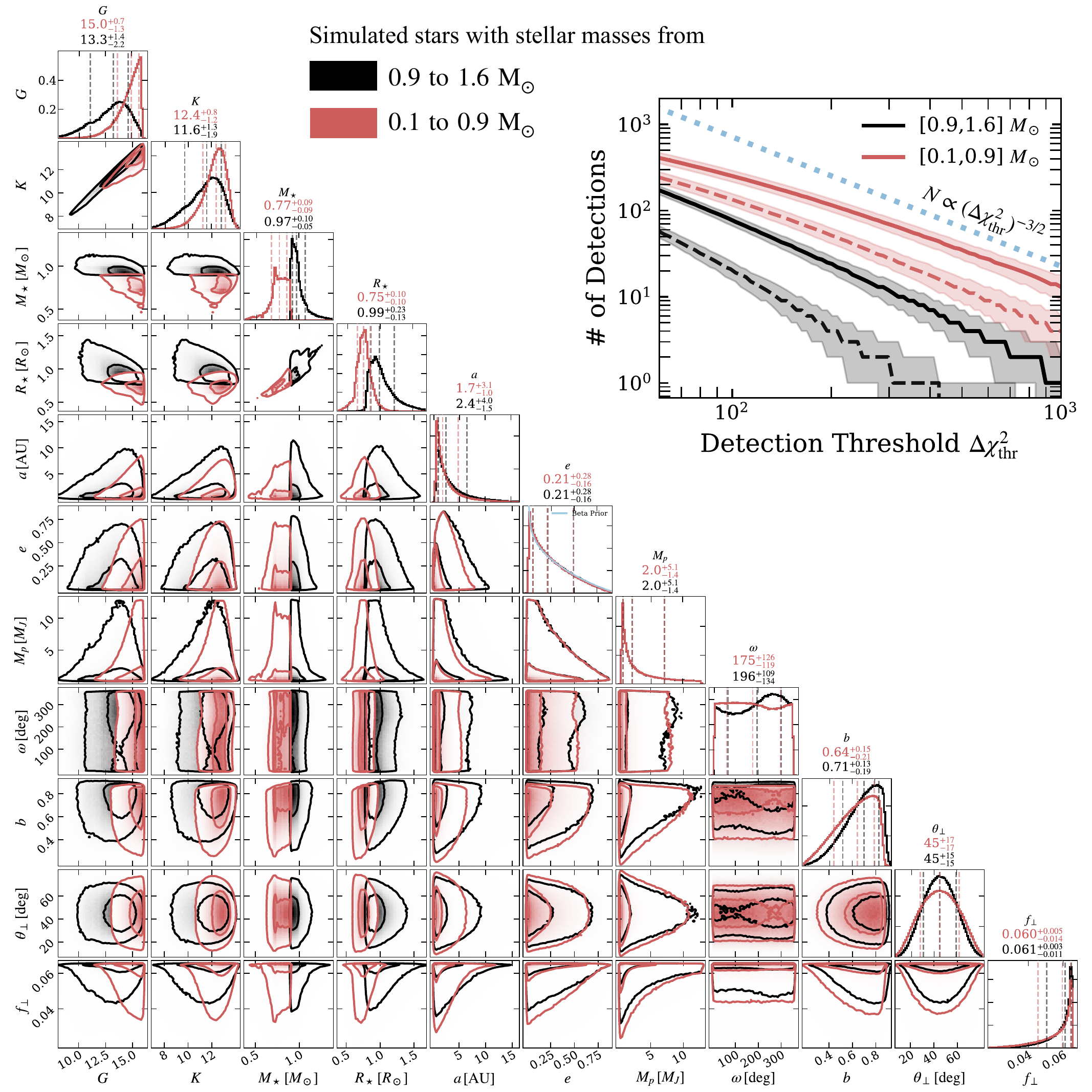}
    \caption{Parameter distributions of simulated systems for
    which Jupiter-like oblateness is detectable with $\Delta\chi^2> 60$,
    assuming an isotropic distribution
    of true obliquities.
    Results are shown separately for
    stellar masses $0.9<M_\star/M_\odot<1.6$ (black)
    and $0.1<M_\star/M_\odot<0.9$ (red).
    Above each histogram are the median and 16th/84th percentiles.
    The upper right panel shows how the number
    of detections scales with the $\Delta\chi^2$ threshold. Solid curves use the ETC noise
    model, dashed curves use the empirical noise model, and shaded bands indicate
    the $1\sigma$ range. The
    dotted line shows the reference scaling
    $N\propto(\Delta\chi^2_{\rm thr})^{-3/2}$.}    
    \label{fig:properties_oblateness}
\end{figure*}

\begin{figure*}[t!]
    \centering
    \includegraphics[width=1.0\linewidth]{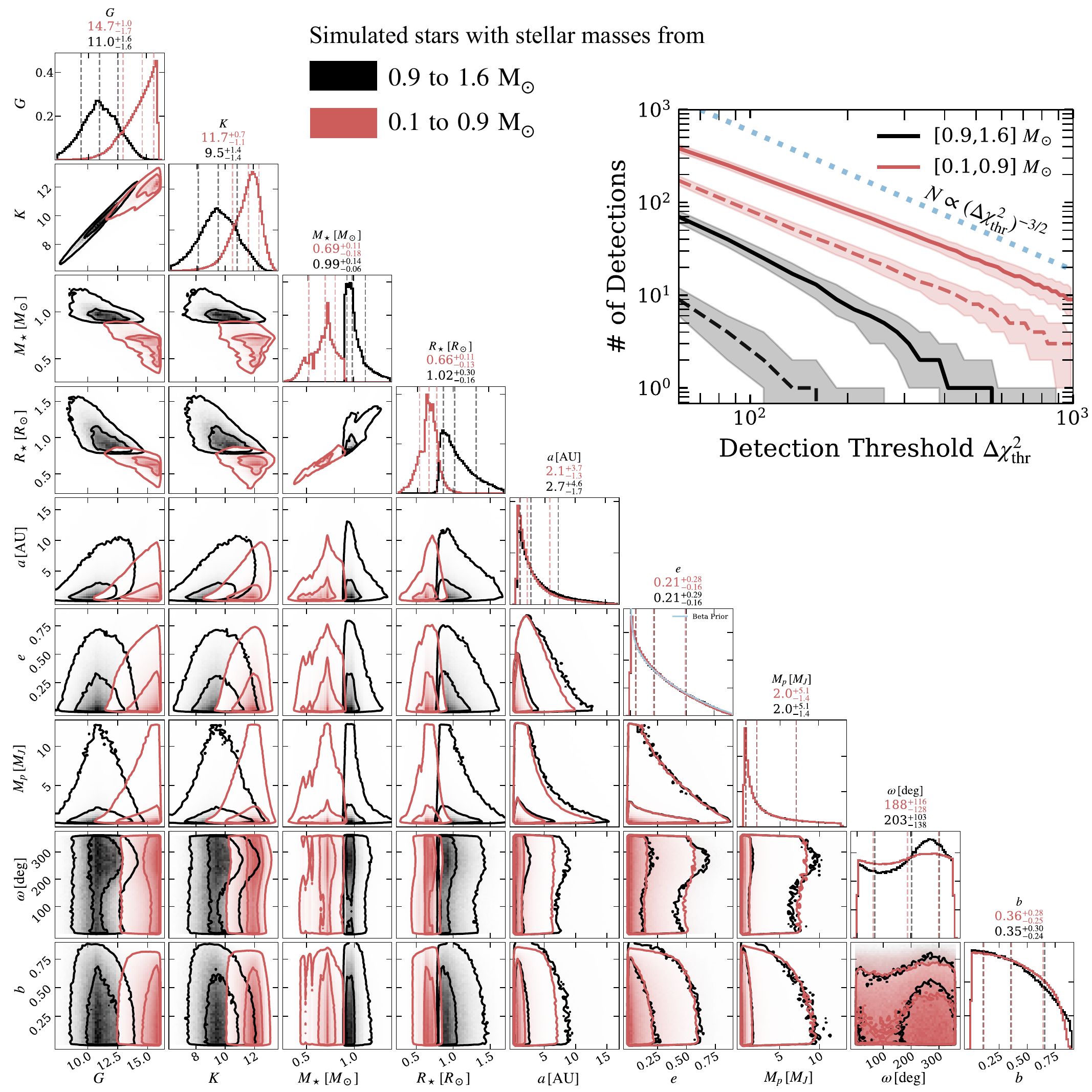}
    \caption{Same as Figure~\ref{fig:properties_oblateness}, but for systems around which a Ganymede-sized
    moon would be detectable if present.}
    \label{fig:properties_moon_full}
\end{figure*}

\begin{figure*}[t!]
    \centering
    \includegraphics[width=1.0\linewidth]{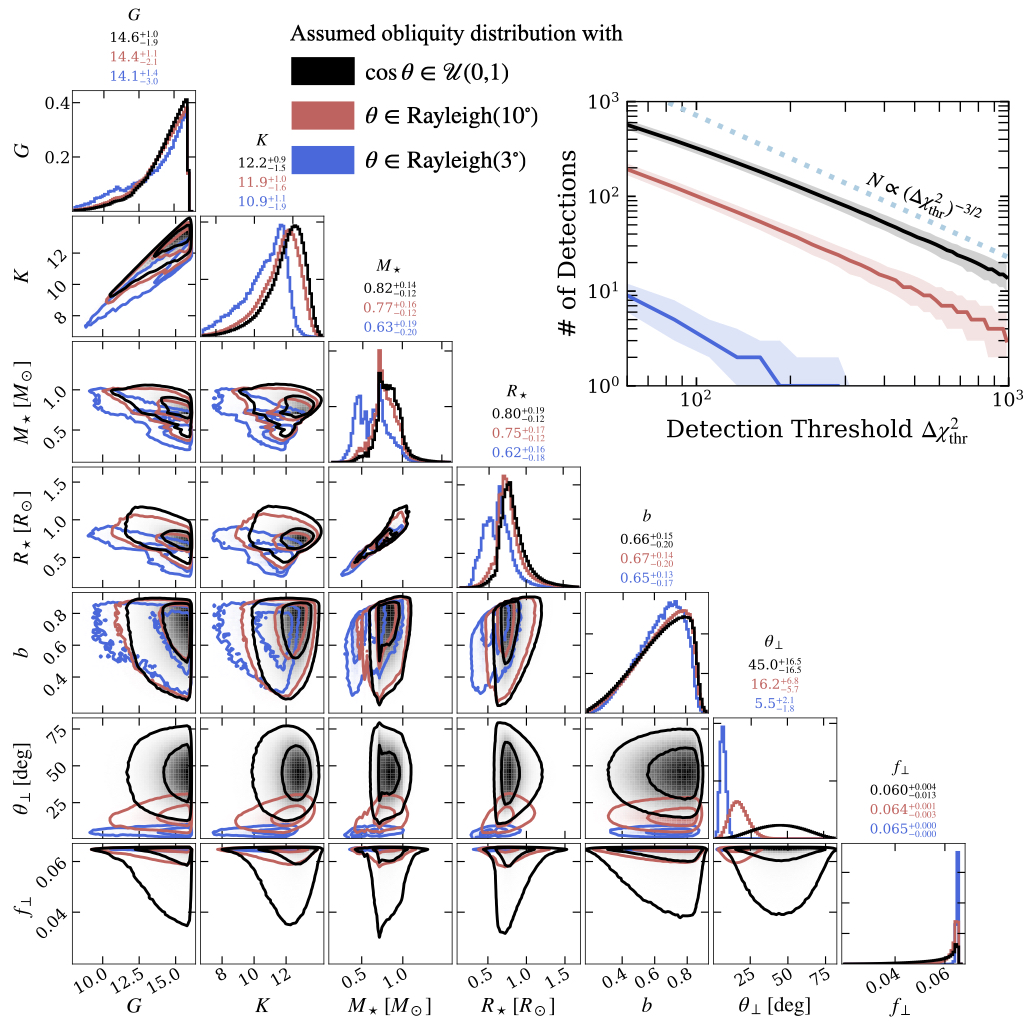}
    \caption{Parameter distributions of simulated systems for
    which Jupiter-like oblateness is detectable with $\Delta\chi^2> 60$, shown
    for three true-obliquity distributions:
    isotropic (black),
    Rayleigh peaking at $10^\circ$ (red),
    and Rayleigh peaking at $3^\circ$ (blue).
    Above each histogram are median values and 16th/84th percentiles.
    The upper-right panel shows how the number
    of detections scales with the $\Delta\chi^2$ threshold for the ETC noise model;
    shaded bands indicate the $1\sigma$ range, and
    the dotted line shows the 
    reference scaling $N\propto(\Delta\chi^2_{\rm thr})^{-3/2}$.}    
    \label{fig:obl_obliquity}
\end{figure*}

\begin{table*}[t]
%\centering
\caption{Predicted numbers of
systems with $\Delta\chi^2>60$.
For oblateness, the table gives
the number of systems in which Jupiter-like
rotational oblateness would be detectable
under the specified obliquity prior.
For moon searches, the table gives the number
of systems for which a Ganymede-sized moon
would be detectable if present.
Results are given for two different
JWST noise models and for two stellar samples:
the baseline sample with $0.9<M_\star/M_\odot<1.6$,
and the lower-mass sample
with $0.1<M_\star/M_\odot<0.9$.
\label{tab:yield_favorable_systems}
}
\begin{tabular}{llccc}
\toprule
Type of Detection & Prior & Stellar sample & ETC & Empirical noise model \\
\midrule
\textbf{Jupiter-like oblateness}
  & $\cos\theta \sim \mathcal{U}(0,1)$
  & $0.9$--$1.6~M_\odot$ & $168^{+19}_{-19}$ & $59^{+10}_{-9}$ \\
  &
  & $0.1$--$0.9~M_\odot$ & $402^{+60}_{-55}$ & $247^{+39}_{-36}$\\

\addlinespace[0.04in]
  & $\theta \sim \mathrm{Rayleigh}(10^\circ)$
& $0.9$--$1.6~M_\odot$ & $40^{+8}_{-7}$ & $10^{+4}_{-3}$ \\
  & 
  & $0.1$--$0.9~M_\odot$ & $153^{+25}_{-24}$ & $79^{+15}_{-14}$ \\

\addlinespace[0.04in]
  & $\theta \sim \mathrm{Rayleigh}(3^\circ)$
  & $0.9$--$1.6~M_\odot$ & $1^{+1}_{-1}$ & $0$ \\
  &
  & $0.1$--$0.9~M_\odot$ & $8^{+3}_{-3}$ & $2^{+2}_{-1}$ \\

\midrule
\textbf{Ganymede-sized moon}
  & --
  & $0.9$--$1.6~M_\odot$ & $70^{+11}_{-10}$ & $9^{+3}_{-3}$ \\
  &
  & $0.1$--$0.9~M_\odot$ & $382^{+39}_{-36}$ & $172^{+20}_{-19}$ \\
\bottomrule
\end{tabular}
\end{table*}

Table~\ref{tab:yield_favorable_systems}
gives the predicted
yields for detections
of Jupiter-like oblateness
and for systems
that can be effectively searched
for Ganymede-sized moons,
using both the ETC-based and the
empirical JWST noise models for single-transit
observations.
The quoted
uncertainties
were derived from 
$10^4$ Monte Carlo realizations
and reflect random orbital
orientations, spin-axis orientations,
assignments of planets to stars,
and uncertainties in the planet-occurrence
model.
For the lower-mass stellar sample,
there is an additional systematic uncertainty
from the poorly known occurrence rate
of giant planets.

The predicted yield of oblateness
detections depends sensitively on the
assumed distribution of true obliquities.
In the baseline stellar sample
(0.9--1.6$M_\odot$),
and using the empirical noise model,
a Rayleigh distribution peaking at $10^\circ$
gives $10^{+4}_{-3}$ detections.
If JWST reaches the ETC-predicted precision,
the yield rises to
$40^{+8}_{-7}$.
If obliquities are isotropically
distributed, the expected yields
are larger by a factor of 4--6.
If they are instead
concentrated near Jupiter's $3^\circ$
obliquity, the yield is consistent
with zero for the baseline stellar sample.

For Ganymede-sized moons,
the numbers in Table~\ref{tab:yield_favorable_systems}
should be interpreted as the
number of systems in which such a moon would
be detectable if present.
In the baseline stellar sample,
the empirical noise model gives
$9^{+3}_{-3}$ favorable systems, while the
ETC-based model gives $70^{+11}_{-10}$.
The actual number of detections would depend on the occurrence
rate and radius distribution of moons, which
are presently unknown.

Including lower-mass host stars substantially
increases all of the predicted yields.
With the empirical noise model,
the lower-mass sample
adds $79^{+15}_{-14}$ detectable systems for the Rayleigh ($10^\circ$) obliquity prior and $172^{+20}_{-19}$ systems favorable
for Ganymede-sized moon searches. These numbers should be regarded as
less robust than the baseline-sample results, because of uncertainties in both
giant-planet occurrence and the ability
of transit surveys to achieve high completeness
at faint
apparent magnitudes.

Figure \ref{fig:properties_oblateness}
and Figure \ref{fig:properties_moon_full}
show the properties of systems
that pass the $\Delta\chi^2 > 60$
threshold for Jupiter-like
oblateness and Ganymede-sized moon
searches, respectively,
using the ETC noise model as an optimistic assumption.
In both cases, the favorable systems
are strongly weighted toward lower-mass stars.
The planets typically
have orbital distances of about
0.8--7~AU, reflecting the competition
between rising planet occurrence and falling
transit probability as orbital distance increases.
As expected, oblateness detections
favor impact parameters $b\approx 0.7$ and
projected obliquities $\theta_\perp \approx 45^\circ$,
while moon-searchable systems
favor lower impact parameters
because they produce longer transit
durations.
Detecting a Ganymede-sized 
moon generally requires a host
star about 1~mag brighter
than is needed to detect
Jupiter-like oblateness.

Figure \ref{fig:obl_obliquity} shows how
the properties of the systems with
detectable oblateness change with
the assumed obliquity distribution.
As the typical obliquities become smaller,
the detectable sample shifts further
towards lower-mass and brighter host stars.
This trend again emphasizes the
value of low-mass stars. Because their
smaller radii amplify the photometric
signal, oblateness can be detected
over a wider range of spin-axis orientations.

The total yield is also sensitive
to the adopted $\Delta\chi^2$ threshold. As shown
in the upper right panels of
Figs.~\ref{fig:properties_oblateness}, \ref{fig:properties_moon_full}, and \ref{fig:obl_obliquity},
the number of systems approximately
follows $N\propto (\Delta\rm \chi^2_{thr})^{-3/2}$,
as expected for
a uniform spatial density of stars
\footnote{At fixed luminosity, photon-limited
precision $\sigma$ is proportional to distance $d$, and $\Delta\chi^2 \propto 
\sigma^{-2} \propto d^{-2}$. Thus, the limiting distance
is proportional to $(\Delta\chi^2)^{-1/2}$ and the
search volume is proportional to $(\Delta\chi^2)^{-3/2}$.}
Increasing the threshold 
from $\Delta\chi^2=60$
to 100 decreases the predicted yields
by a nearly 
a factor of three.
The assumed JWST noise model
is also crucial:
using the empirical noise model
instead of the ETC-based model 
lowers the yield by factors of a few
to nearly an order of magnitude.
The strongest detections are expected
to have $\Delta\chi^2$ values
between 100 and 1000, depending on
the stellar sample and noise model.

\section{Sensitivity to Correlated Noise}\label{sec:rednoise}

\begin{figure}[b!]
    \centering
    \includegraphics[width=1.0\linewidth]{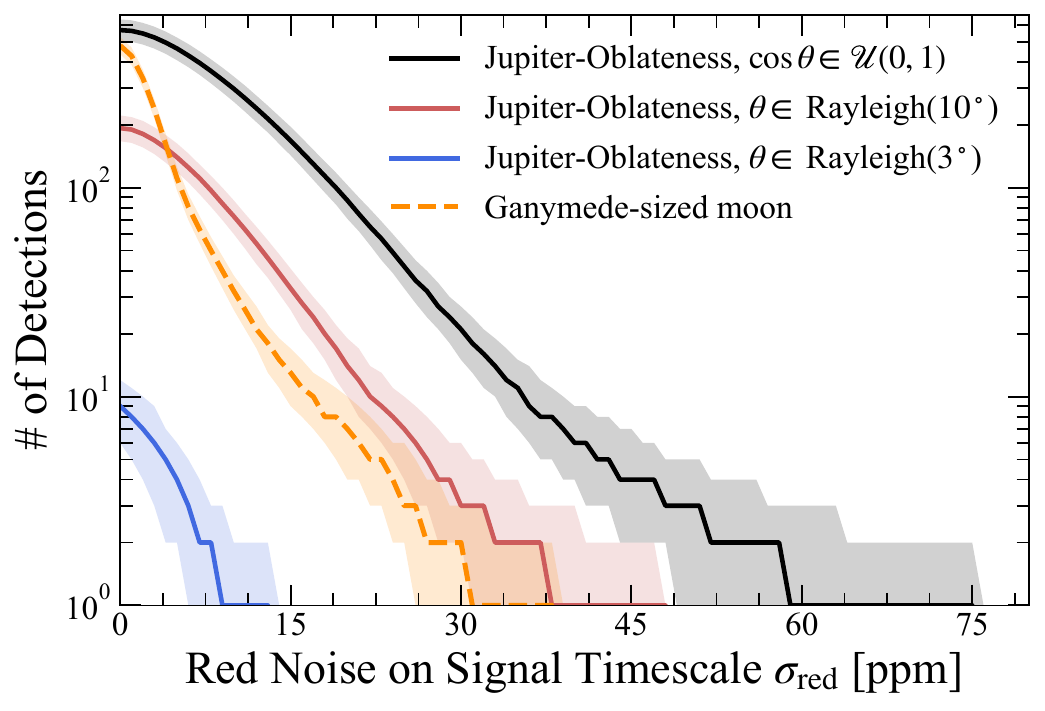}
    \caption{Sensitivity of the predicted yields to
    an added red-noise floor 
    on the relevant signal timescale.
    The horizontal axis specifies $\sigma_{\rm red}$, the
    additional noise added in quadrature to the ETC
    white-noise
    prediction after averaging to
    the relevant timescale (half the transit duration
    for moon detection, and half the ingress/egress duration
    for oblateness detection).
    Solid curves show the number of systems with
    detectable Jupiter-like oblateness for the three
    assumed obliquity distributions.
    The dashed orange curve shows the number of systems for
    which a Ganymede-sized moon would be detectable.
    Shaded regions show the 16th--84th percentile range over Monte Carlo realizations.
    In all cases, the host
    star sample was the combined 0.1--1.6~$M_{\odot}$ sample.}
    \label{fig:red_noise}
\end{figure}

To this point, our forecasts assume that the photometric precision
is accurately described by white-noise models, in which the uncertainties of the data points are uncorrelated.
The models predict the 1-min precision as a function of apparent magnitude and assume that time-averaging $N$ data points would
reduce the fluctuation level by a factor of $\sqrt{N}$.
However, in real light curves, the actual noise
level on longer timescales can be larger than expected
from white-noise averaging. The sources of such ``red noise''
include instrumental systematics, imperfect detrending, and
stellar variability.

Because the amplitude and timescale dependence of red noise
depend on the details of the instrument, data processing,
and stellar type, we do not attempt to build a universal
red-noise model. Instead, we ask how the yields change
if there is an irreducible noise floor on the relevant
signal timescale, which we take to be
half of the total transit duration for moon detection,
and half of the ingress/egress duration for oblateness
detection. We model the noise on the relevant timescale as
\begin{equation}
\label{eq:red_noise}
\sigma^2
=
\frac{\sigma_{{\rm ETC}}^2}{N}
+
\sigma_{{\rm red}}^2,
\end{equation}
where $\sigma_{\rm ETC}$ is the 1-minute noise level
from Equation~\ref{eq:noise_etc},
$N$ is the number of minutes in the relevant timescale,
and $\sigma_{\rm red}$ is a constant representing
the noise floor on that timescale. We use the ETC model
as the white-noise reference case, rather than the empirical
noise model, to better isolate the effect of the added red-noise
term.

%\citep[see, e.g.,][]{Kipping2025,BelloArufe2025, cassese2026}. 

Figure~\ref{fig:red_noise} shows that the predicted yields are sensitive to the red-noise floor.
The oblateness yield becomes consistent with zero at
$\sigma_{{\rm red}}\sim60$, $35$, and $10~{\rm ppm}$
for the isotropic, Rayleigh $10^\circ$, and Rayleigh~$3^\circ$
obliquity priors, respectively. The Ganymede-sized moon yield
is also strongly affected, becoming consistent with zero for
$\sigma_{{\rm red}}\sim30~{\rm ppm}$. Thus, correlated noise
at the level of a few tens of ppm on timescales of 1--10~hr
can eliminate many systems that would otherwise appear
favorable under the white-noise assumption.

% the former two studies examined oblateness and Ganymede-sized moons in Kepler-167 e and the latter study examined the lunar-sized moon around TOI-700 d and e.
Recent JWST searches illustrate the importance of this issue.
For Kepler-167\,e, the half-ingress/egress duration is approximately an hour. Using Equation~\ref{eq:moon_empirical_etc_used},
the precision required to reach $\Delta\chi^2=60$ for Jupiter-like
oblateness at an optimal sky projection is
$\sigma\simeq 15$~ppm.
The observed precision was instead about 30~ppm on this timescale (Figure 7 of \citealt{cassese2026}).
Similarly, for lunar-sized moons around TOI-700\,d and e, 
the half-transit durations are 99 and 92 minutes, and the required precision on that timescale is about 6.5~ppm.
The observed residuals on comparable timescales were instead
about 16 and 20~ppm (Figure 5 of \citealt{Pass2026}).
In each case, the noise on the relevant signal timescale exceeded
the precision needed for a secure detection.

This exercise shows that the yields in Section~\ref{sec:detectable_systems} should be interpreted
as optimistic forecasts that apply when time-correlated noise is
small. They remain useful for identifying the regions of parameter
space where JWST could be sensitive to oblateness or moons,
but the feasibility of any particular observation
must also be assessed using the measured or expected noise
on the appropriate timescale for the target.

% The presence of correlated noise is the key reason that prevented definitive detection of oblateness and moons by \cite{Kipping2025}, \cite{cassese2026}, and \cite{Pass2026}; the former two studies searched for oblateness and Ganymede-sized moons in Kepler-167 e, and the latter group examined the lunar-sized moon around TOI-700 d and e. To see this, we can derive the photometric precision required to detect oblateness and moons by rearranging Eqns \ref{eq:oblateness_empirical_etc_app} and \ref{eq:moon_empirical_etc_used}:
% \begin{equation} 
% \frac{\Delta\chi^2_{\rm obl}}{60} \simeq F^2 \left( \frac{16.6~{\rm ppm}}{\sigma_{\tau/2}} \right)^2 \left( \frac{\delta}{\delta_{J\odot}} \right)^2 , \label{eq:oblateness_tauhalf_precision} 
% \end{equation} and 
% \begin{equation} 
% \frac{\Delta\chi^2_{\rm moon}}{60} \simeq \left( \frac{2.52~{\rm ppm}}{\sigma_{T/2}} \right)^2 \left( \frac{\delta_m}{\delta_{G\odot}} \right)^2. 
% \label{eq:moon_Thalf_precision} 
% \end{equation} 
% Here $\delta_{J\odot}=(R_J/R_\odot)^2$, and $\delta_{G\odot}=(R_{\rm Ganymede}/R_\odot)^2$.}

\section{Discussion}\label{sec:discussion}

\subsection{Limitations of the forecast}

Our forecasts are intended to identify the systems
and regions of parameter space where JWST could enable the
detection of moons and rotational oblateness of giant
exoplanets. Several factors could change the numerical yields,
some of which are related to the detectability of a given signal, and some of which are related to unknown astrophysical
parameters.
The sensitivity
to correlated noise was discussed in Section~\ref{sec:rednoise}.
The main astrophysical uncertainties arise from
the unknown obliquity distribution of giant planets
and the unknown occurrence rate and size distribution
of large moons -- indeed, reducing these uncertainties
is one of the main reasons for attempting such observations.
If giant-planet obliquities are typically as low as Jupiter's,
oblateness detections will be rare; if they are comparable
to Saturn's or even more broadly distributed, the yield
could be much higher. Likewise, the actual number of moon
detections could be smaller or larger than the expected
number of moon-searchable systems depending on the abundance
of large moons and systems of multiple moons.

% First, our forecast for moon detections implicitly assumes one moon per planet.  This assumption appears to be at least broadly consistent with the Solar System, where each giant planet hosts multiple moons. In reality, the occurrence rate and size distribution of large moons around giant exoplanets are currently unknown. The true yield could therefore be either lower or higher than our estimate, depending on how common such moons actually are. Similarly, the planet obliquity distribution is unknown. As we showed in Table \ref{tab:yield_favorable_systems}, orders-of-magnitude oblateness yield difference could occur given different obliquity distributions.

Uncertainties in stellar and planetary
properties also propagate directly into the
forecasts. The detectability metrics depend sensitively on stellar radius, and errors in $R_\star$ can bias the results.
The radii and masses from the Gaia DR3 astrophysical-parameters table have
been found to be generally reliable \citep{creevey2023, Fouesneau2023}, but
residual uncertainties
and extinction-related systematics may
be important,
especially at larger distances.

The moon calculation was idealized,
assuming that the moon produces a distinct, non-overlapping transit.
Real planet-moon light curves can be more
complex, and degeneracies may exist
with planet model parameters. 
Multiple transit observations would help to
break these degeneracies because
the planetary transit is
repeatable whereas the moon transit
changes in timing and duration.

\begin{figure}[t!]
    \centering
    \includegraphics[width=1.0\linewidth]{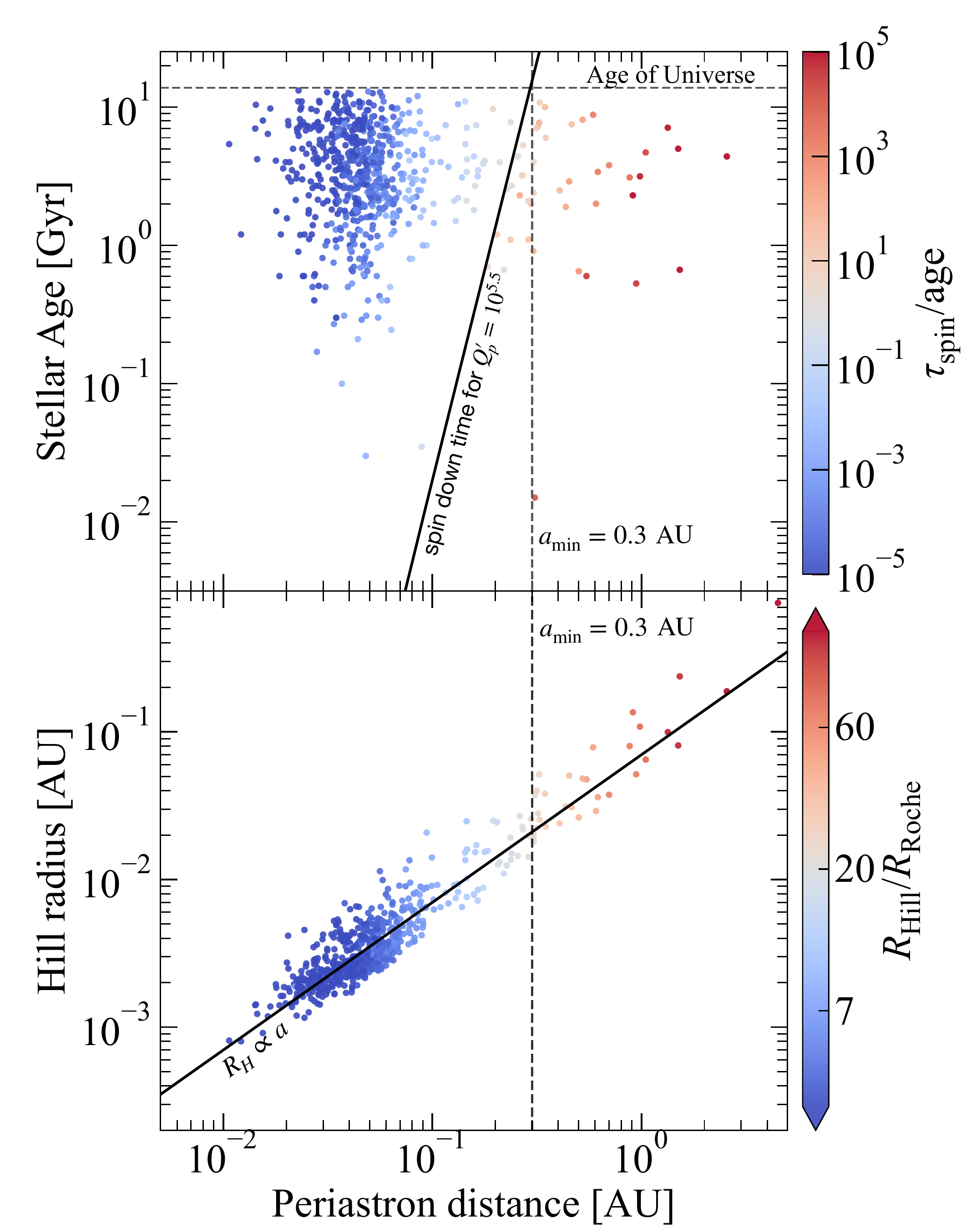}
    \caption{Effects of relaxing the 0.3~AU periastron distance. \textit{Top row:} Tidal despinning timescales
     for known transiting giant planets.
     Points show stellar age versus periastron distance
     and are colored according to $\tau_{\rm spin}$/Age, where
     $\tau_{\rm spin}$ is evaluated system-by-system assuming $Q_p' = 10^{5.5}$. Solid curves show $\tau_{\rm spin}$ for $Q_p' = 10^{5.5}$, assuming Jupiter-like planetary properties. The dashed horizontal line marks the age of the universe, and the dashed vertical line marks the adopted 0.3~AU cut. \textit{Bottom row:} Hill-sphere
     constraints for moons. Points show the planetary
     Hill radius as a function of periastron
     distance and are colored according to $R_{\rm Hill}/R_{\rm Roche}$, with the Roche distance evaluated assuming a satellite density of $3~\rm g/cm^{-3}$. The color scale is centered at $R_{\rm Hill}/R_{\rm Roche} = 20$, motivated by the fact that most large regular satellites in the Solar System orbit within $0.05~R_{\rm Hill}$. The adopted 0.3~AU cut is conservative: some systems at smaller periastron distances may retain measurable rotational oblateness or have large enough Hill radii for long-lived moons.}
    \label{fig:spindown}
\end{figure}

Finally, we imposed a conservative
0.3~AU minimum periastron distance
to exclude systems likely to have undergone
tidal despinning and to favor systems
with adequate room within the Hill radius
for long-lived moons.
However, there should be no sharp
boundary in reality.
Some younger systems or systems
with higher $Q_p'$ values
may retain rapid rotation even at smaller separations,
and some closer-in planets
may still have stable moons.
Figure \ref{fig:spindown}
illustrates how relaxing this cut
could increase the
number of favorable systems.

% the (somewhat arbitrary) factor of 0.05 is chosen conservatively based on the distance for the majority of the Solar System satellites and 

The calculations assumed fiducial
Jupiter-like planets and Ganymede-sized moons.
The scaling relations make it straightforward
to estimate the effects on the yield of
other choices.
For oblateness,
the yield scales approximately
as $(f^2 R_p^5)^{3/2}$. Thus,
replacing Jupiter-like planets
with Saturn-like
planets ($R_{\rm p} = 0.84_{\rm J}$, $f=0.098$)
changes the yield by a factor
\begin{equation}
\frac{N_{\rm Sat}}{N_{\rm Jup}} =
\left[\left( \frac{0.098}{0.065} \right)^{\!\!2}
\left( \frac{0.84}{1.00} \right)^{\!\!5}\right]^{3/2} \approx 1.0,
\end{equation}
i.e., the smaller radius of Saturn is compensated
almost exactly by its higher oblateness.

For moons, the yield
scales approximately as
$(R_m^4)^{3/2} = R_m^6$.
Thus, replacing Ganymede-sized moons
with Earth-sized moons
increases the expected number of moon-searchable
systems by a factor
\begin{equation}
\frac{N_{\rm \oplus}}{N_{\rm G}} =
\left( \frac{R_\oplus}{R_{\rm G}}  \right)^{\!\!6}
\approx 200.
\end{equation}

\subsection{Comparison with Known Systems}

\begin{figure*}[t!]
    \centering
    \includegraphics[width=1.0\linewidth]{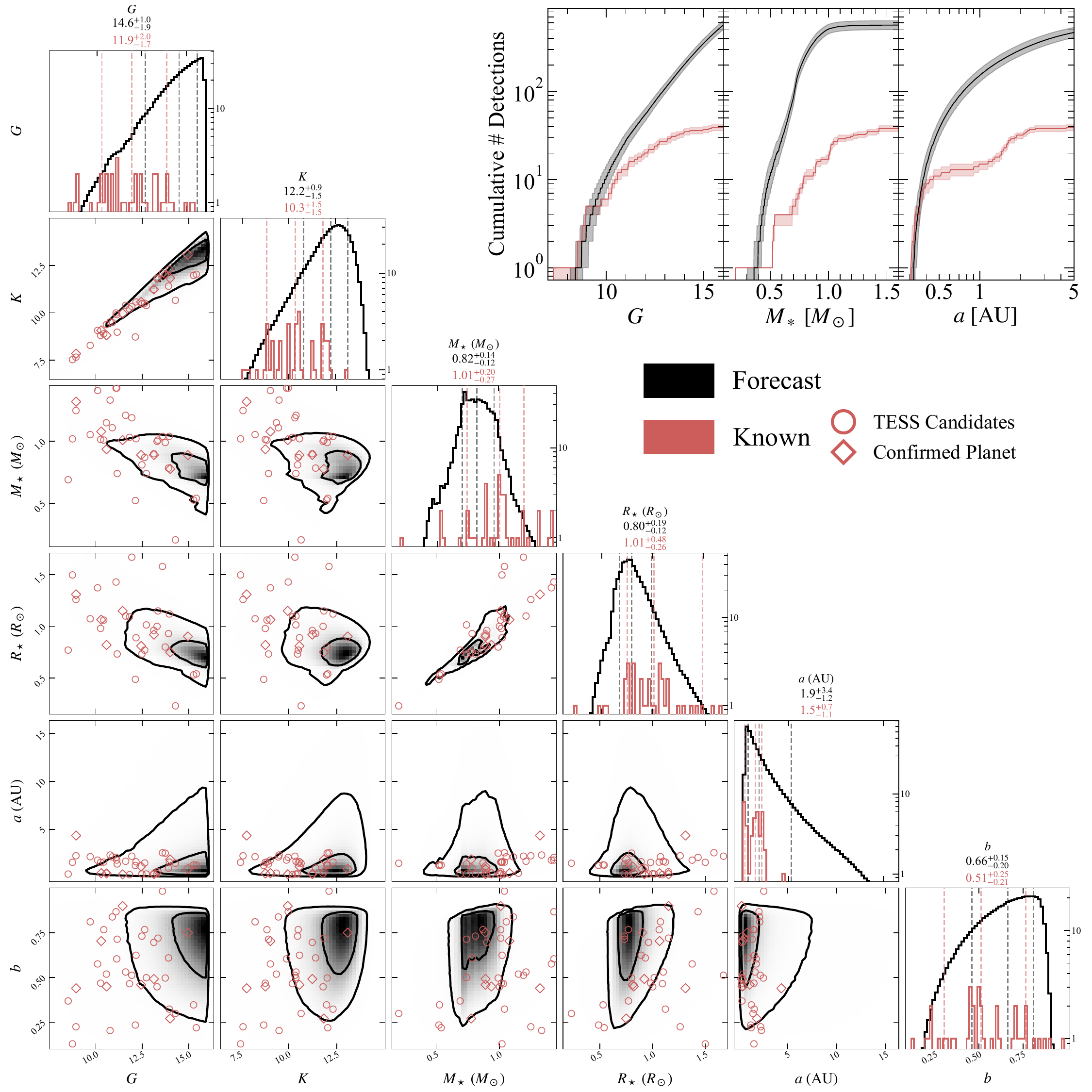}
    \caption{ Corner plot comparing the forecasted oblateness-favorable
    population (black) with the currently known confirmed and candidate
    systems (red). 
    The upper-right panel
    shows cumulative distributions of $G$, $M_\star$, and $a$.
    The observed sample is fairly complete at the brightest magnitudes and
    smallest
    separations, but shows a growing deficit toward optically fainter
    stars, lower-mass stars, and larger orbital separations.
    } 
    \label{fig:comp_real_obl}
\end{figure*}

\begin{figure*}[t!]
    \centering
    \includegraphics[width=1.0\linewidth]{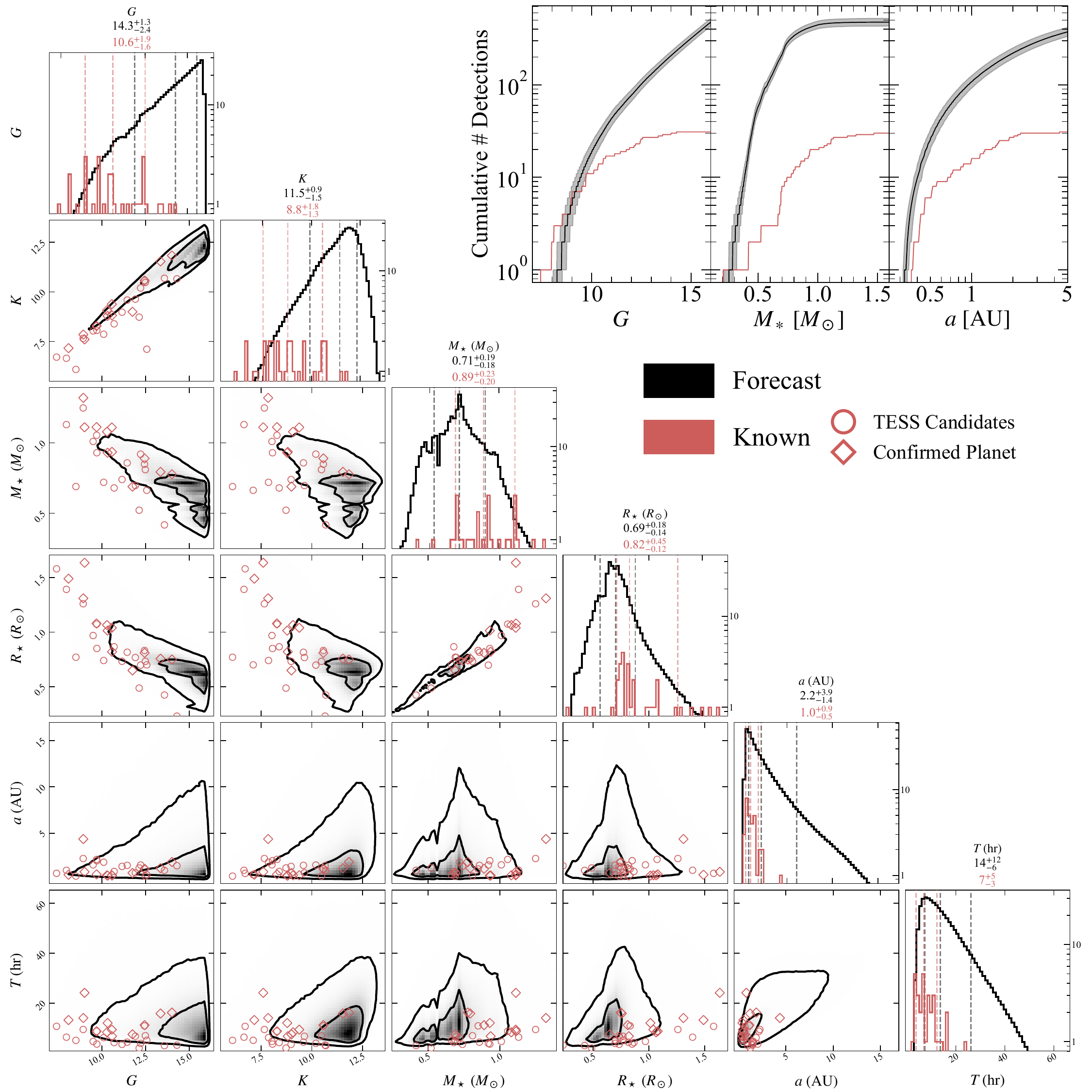}
    \caption{Same as Figure~\ref{fig:comp_real_obl}, but
    for systems that are searchable for Ganymede-sized moons.}
    \label{fig:comp_real_moon}
\end{figure*}

A natural question is whether the 
favorable systems predicted by our model are already
present in the current catalog of transiting planets.
To address this question, we evaluated the same detectability metrics
for confirmed planets in the NASA Exoplanet Archive\footnote{\href{https://exoplanetarchive.ipac.caltech.edu}{https://exoplanetarchive.ipac.caltech.edu}} \citep{exoarchive},
restricting the sample
to planets with $R_{\rm p}> 0.33\,R_{\rm J}$ and
periastron distances larger than 0.3~AU. We also
evaluated the metrics for TESS planet candidates.\footnote{For TESS
candidates, impact parameters and eccentricities are usually
unavailable; we assumed circular orbits and drew $b$ from $U(0,1)$.}

Using a threshold of $\Delta\chi^2> 60$ and the ETC-based noise model,
there are 8 confirmed systems that are favorable for detecting Jupiter-like
oblateness and 12 that are searchable for Ganymede-sized moons.
Including the list of TESS planet candidates
raises these numbers to 39 and 31. 
These counts are
well below the $\sim$100 favorable
systems predicted by our forecast at
similar brightness levels ($G\lesssim 12.5$), suggesting
that the current transiting-planet catalog remains
incomplete for this application.

Figure \ref{fig:comp_real_obl} and Figure \ref{fig:comp_real_moon} compare
the forecasted favorable populations with the known confirmed
and candidate systems. The currently known favorable systems are concentrated around bright Sun-like stars and at shorter orbital
periods, reflecting the selection function of existing transit surveys.
For oblateness searches, the sample appears to be nearly complete
for $G\lesssim 10$. For Ganymede-sized moon searches,
completeness appears high for $G\lesssim 9.5$.
At fainter magnitudes, lower stellar masses, and larger orbital
distances, the observed sample falls increasingly short
of the forecast. The deficit is especially
pronounced for $M_\star < 0.8\,M_\odot$
and $a\gtrsim 0.4$~AU.

This pattern is consistent with known selection effects.
TESS and most other transit surveys
favor short-period planets around optically bright stars,
whereas the systems most favorable for oblateness and moon searches
tend to have longer periods and, often, lower-mass host stars.
Thus, the apparent absence of favorable systems is
best explained as the result of survey incompleteness in the relevant
region of parameter space, rather than
intrinsic scarcity.

\subsubsection{The Best Known Systems}

\begin{table*}[t!]
\centering
\begin{threeparttable}
\caption{Confirmed systems offering best
prospects for detecting oblateness and
moons.}
\label{tab:favorable_confirmed_systems}

\begin{tabular}{lcccccccccc}
\toprule
Planet & $M_\star$ ($M_\odot$) & $R_\star$ ($R_\odot$) & $K$ & $G$ & $R_p$ ($R_J$) & $b$ & $a$ (AU) & $e$ & $\Delta\chi^2_{\rm ETC}$ & $\Delta\chi^2_{\rm emp}$ \\
\midrule
\multicolumn{11}{c}{\textbf{Oblateness}} \\
\midrule
TOI-2449 b     & 1.08 & 1.07 &  9.07 & 10.29 & 1.00 & 0.704 & 0.45 & 0.10 & 429 & 123 \\
TOI-199 b      & 0.94 & 0.82 &  8.81 & 10.58 & 0.81 & 0.450 & 0.43 & 0.09 & 247 &  72 \\
TOI-201 c\tnote{$\dagger$}      & 1.32 & 1.34 &  7.85 &  8.94 & 0.99 & 0.440 & 4.28 & 0.64 & 210 &  49\\
TIC 4672985 b  & 1.01 & 1.15 &  9.97 & 11.44 & 1.03 & 0.900 & 0.33 & 0.02 &  147 &  50 \\
TOI-4600 c   & 0.89 & 0.81 & 10.57 & 12.44 & 0.84 & 0.460 & 1.15 & 0.21 &  130 &  57 \\
Kepler-553 c   & 0.89 & 0.90 & 13.06 & 14.96 & 1.03 & 0.751 & 0.90 & 0.35 &  119 &  75 \\
TOI-4504 c     & 0.89 & 0.92 & 11.24 & 13.13 & 0.99 & 0.492 & 0.36 & 0.03 &  78 &  44 \\
Kepler-167 e   & 0.78 & 0.75 & 11.83 & 14.00 & 0.91 & 0.271 & 1.88 & 0.29 &  71 &  41 \\
\midrule
\multicolumn{11}{c}{\textbf{Moons}}\\
\midrule
Kepler-16 b\tnote{$\mathsection$}    & 0.69 & 0.65 & 9.00  & 11.75 & 0.75 & 0.132 & 0.70  & 0.01 & 418 & 116 \\
TOI-199 b      & 0.94 & 0.82 &  8.81 & 10.58 & 0.81 & 0.450 & 0.43 & 0.09 & 163 &  46 \\
TOI-4465 b     & 0.93 & 1.01 &  8.79 & 10.33 & 1.25 & 0.080 & 0.42 & 0.24 & 130 &  36 \\
TOI-2010 b     & 1.11 & 1.08 &  8.28 &  9.70 & 1.05 & 0.147 & 0.55 & 0.21 & 102 &  29 \\
TOI-201 c\tnote{$\dagger$}      & 1.32 & 1.34 &  7.85 &  8.94 & 0.99 & 0.440 & 4.28 & 0.64 &  90 &  28 \\
TOI-4600 c    & 0.89 & 0.81 & 10.57 & 12.44 & 0.84 & 0.460 & 1.15 & 0.21 &  80 &  34 \\
HD 114082 b & 1.47 & 1.49 & 7.16 & 8.09 & 1.00 & 0.000 & 0.51 & 0.40 & 80 & 24\\ 
TOI-2180 b     & 1.11 & 1.64 &  7.60 &  9.01 & 1.01 & 0.100 & 0.83 & 0.37 &  80 &  23 \\
Kepler-421 b   & 0.79 & 0.76 & 11.54 & 13.35 & 0.37 & 0.210 & 1.22 & 0.04 &  76 &  42 \\
Kepler-167 e   & 0.78 & 0.75 & 11.83 & 14.00 & 0.91 & 0.271 & 1.88 & 0.29 &  69 &  39 \\
TOI-4507 b     & 1.11 & 1.04 &  9.38 & 10.57 & 0.73 & 0.120 & 0.46 & 0.09 &  61 &  17 \\
TOI-2449 b     & 1.08 & 1.07 &  9.07 & 10.29 & 1.00 & 0.704 & 0.45 & 0.10 &  61 &  17 \\
\bottomrule
\end{tabular}
\begin{tablenotes}[flushleft]
\footnotesize
\item[$\dagger$] At present, the ephemerides of TOI-201\,c are too uncertain
for scheduling to be feasible. There is also evidence that TOI-201\,c could be a brown dwarf \citep{Mireles2026}.
%was discovered through transit timing variation and has an approximately 8 yr orbit; its ephemeris is uncertain at the roughly year level.
\item[$\mathsection$] Kepler-16\,b is a special case, as a circumbinary planet. It has precessed out of the line-of-sight and will not be transiting until 2042 \citep{Doyle2019}.
\end{tablenotes}
\end{threeparttable}
\end{table*}

Table~\ref{tab:favorable_confirmed_systems} lists the most favorable confirmed planets.
The leading targets are predominantly TESS-discovered
systems. Five systems (TOI-199\,b, TOI-2449\,b, TOI-201\,c, TOI-4600\,c, and Kepler-167\,e) appear in both the oblateness and moon-search lists.
They are therefore especially valuable targets for JWST, provided
their ephemerides are sufficiently accurate, the transit observations
can be scheduled, and stellar variability does
not produce incorrigible noise at a level exceeding tens of ppm
on the relevant timescales.

The ranking helps to explain some of the results of previous
JWST searches.
The super-puff Kepler-51\,d, for which null results
were reported by \citet{lammerswinn2024},
is not expected to yield an oblateness detection in our framework:
$\Delta\chi^2=3.4$ for the ETC noise model, mainly because
of its suboptimal impact parameter ($b = 0.25$) and
relatively faint host star ($K = 13.2$).
Kepler-167\,e is more favorable, with predicted
$\Delta\chi^2 = 71$ for Jupiter-like oblateness and
69 for a Ganymede-sized moon. However, as explained in Section \ref{sec:rednoise}, the actual noise level
was larger than the ETC white-noise expectation,
%\footnote{\textbf{The unexpected systematics were likely caused by the detector resets each time when the on-board software data volume limit was hit, over the course of the 60-hour transit observation.}}
substantially reducing the sensitivity of the JWST observation
\citep{Kipping2025, cassese2026}.
Table~\ref{tab:favorable_confirmed_systems} also demonstrates
that the most intensively studied systems to date
are not the best possible targets.

Our ranking can be compared with the systems identified
by \cite{dholakia2025} as being favorable for oblateness detection.
They used injection-recovery tests with simulated light curves instead
of simple detectability metrics, and 
focused on NIRISS/SOSS observations rather than
selecting an optimal JWST observing mode.
Despite these differences, the results
are broadly consistent.
TOI-199\,b and TOI-4600\,c are found to be 
favorable in both studies.
%Three JWST transit observations have been scheduled to observe TOI-199 b (GO: 5177, P.I. Hu, R; GO: 7188, P.I. Acuna, L.), although one transit showed unexpected systematics at egress that hindered oblateness characterization \citep{BelloArufe2025}.
The targets NGTS-30\,b, TOI-2589\,b, and TOI-1670\,c
were highlighted by \cite{dholakia2025} and are also favorable
by our metrics except that they have periastron distances
smaller than 0.3\,AU and are potentially affected by 
tidal despinning.
Conversely, TOI-2449\,b and TOI-201\,c appear in our list
but were not identified by \cite{dholakia2025}, probably because
they had not yet been confirmed.

Several of these systems have already being observed or are scheduled to
be observed with JWST. TOI-199\,b is the target of approved observations in GO programs 5177 and 7188, although one observed transit showed
unexpected egress systematics that
complicated oblateness characterization \citep{BelloArufe2025}.
HIP~41378\,f, which is not included in our ranked list
because of uncertain
orbital parameters \citep{Garcia2026}, is also
scheduled for JWST observation (GO program 11253). %, P.I. Cassese, B.).

    % \item Kepler-16 b and TOI-2010 b fall just below our conservative detectability threshold of 60, but remain compelling targets because their expected signals are still large ($\Delta\chi^2=55$ and 43, respectively).
    % \item Kepler-553 c was probably excluded by the brightness cut at $J=13$ from \citep{dholakia2025}, where NIRISS/SOSS mode would yield high SNR.
    % \item The absence of TIC 4672985 b from the list reported by \cite{dholakia2025} is less clear. One possibility is that its nearly grazing geometry makes the projected oblateness and obliquity difficult to constrain, even though the expected signal is large.

\section{Summary and Conclusions}\label{sec:conclusion}

We developed an analytic framework to forecast the detectability of planetary oblateness and moons in JWST transit photometry.
We derived scaling relations linking the expected $\Delta\chi^2$ to stellar and planetary properties, orbital geometry, and photometric precision, and validated these relations with injection–recovery simulations.
We then combined these metrics with the Gaia DR3 star catalog
and a giant-planet occurrence model to forecast
the number and properties of the most favorable systems.

The results show that JWST can already carry out meaningful
searches for these signals. Several known systems offer
favorable prospects for detecting Jupiter-like oblateness and searching for Ganymede-sized moons. For our baseline stellar sample
(0.9--1.6~$M_\odot$) and a model based on demonstrated
JWST performance and white-noise assumptions,
the forecasted yields are of order 10 systems for Jupiter-like
oblateness if the true obliquities of giant planets
are typically $\sim$10$^\circ$, and of order 10 systems
in which a Ganymede-sized moon would be detectable if present.
The oblateness yield falls to nearly zero if giant-planet obliquities
are typically as small as Jupiter's 3$^\circ$ obliquity.
Broader obliquity distributions, lower-mass host stars,
and photometric precision closer to the photon-noise limit
can increase the yields to tens or hundreds of systems.
 On the other hand, time-correlated noise of a few tens
of ppm on 1--10~hr timescales can sharply reduce
the predicted yield. Thus, the numbers quoted above should
be considered as favorable-case yields contingent on
the white-noise idealization. 

The current lack of detections does not by itself imply
that these signals are intrinsically rare or beyond JWST's capabilities.
A more immediate limitation is that the most favorable systems
tend to have longer orbital periods, where TESS and other transit
surveys are highly incomplete. Transit surveys are
inherently and strongly biased toward short-period planets,
and therefore undersample the region of parameter space
where rapid planetary spin and long-lived moons are most plausible.

This limitation may be temporary.
The TESS mission has adjusted its sky-scanning
pattern to observe larger fractions of the sky for
longer intervals of continuous coverage.
Ground-based efforts such as
NGTS, ASTEP, WINE, and other collaborations
are increasingly focused on confirming
longer-period transiting planets
\citep[see, e.g.,][]{Wheatley2018, hobson2021, Brahm2023, battley2024, ulmer-moll2025, 
 talapinto2025, Rea2025, kendall2025, robel2026},
and the newly commissioned all-southern-sky HAT-PI observatory 
%(Bakos et al., 2026, in preparation)
may contribute to this effort.
The forthcoming PLATO and Earth 2.0
missions \citep{Rauer2014, Ge+2024} will search large areas of the sky
with space-based precision and longer
continuous baselines.
Meanwhile, the end-of-mission Gaia data release is
to yield tens of thousands of astrometric giant-planet detections, of which $\sim$$10^2$ are expected to transit \citep{lammerswinn2026}.
Finding and confirming this small transiting subset
will require sustained effort, but the planets will occupy
2--5 AU orbits where many favorable systems are expected.

The outlook is therefore promising but conditional.
JWST may be able to measure planetary obliquities,
test whether giant planets
commonly retain rapid rotation, and search for moons, allowing the exoplanet population to be probed for
some of the most familiar aspects of Solar System planets.
The first detections might be close at hand.
To improve the chances of success will require both
a more complete census long-period transiting giant
planets, and methods for target selection
and data analysis to maximize photometric precision
on the relevant signal timescales.
If these challenges can be met, transit photometry
will begin to reveal not only the sizes and atmospheres
of giant planets, but also their spin directions, shapes,
and satellite systems.

\begin{acknowledgments}
We are grateful to Caleb Lammers for perceptive comments on the paper. L.-C.W.\ thanks David Kipping for useful discussions. 
J.N.W.\ thanks Qianyi Shen, Xingjie Zhao, and Haoxu Feng for performing exploratory work on injection-recovery simulations of oblateness detectability that helped motivate this study.
The work reported in this paper was performed using the Princeton Research Computing resources at Princeton University, a consortium led by the Princeton Institute for Computational Science and Engineering (PICSciE) and the Office of Information Technology’s Research Computing. 
\end{acknowledgments}

\appendix

\setcounter{table}{0} % Reset counters
\renewcommand{\thetable}{A\arabic{table}}
\renewcommand\theHtable{Appendix.\thetable}

\setcounter{figure}{0}
\renewcommand{\thefigure}{A\arabic{figure}}
\renewcommand\theHfigure{Appendix.\thefigure}
\section{Geometric Factor}\label{app:G-factor-derivation}

Here we derive the leading-order angular dependence of the
ingress--egress asymmetry produced by 
an oblate planet. In the small-planet limit,
the stellar limb can be treated locally as a straight line,
apart from very central or nearly grazing transits.
The rate at which the planet reduces the starlight
during ingress or egress is proportional to the length of the chord
cut through the planetary shadow by the stellar limb
(Figure~\ref{fig:geometry}). 

Consider an ellipse with semi-major axis $A$ and semi-minor axis $B$.
The maximum chord length cut by a line whose normal makes an angle $\phi$ with the major axis is
\begin{align}
    L = \frac{2AB}{\sqrt{A^2 \cos^2\phi + B^2 \sin^2\phi}}.
    \label{eq:Lmax_f}
\end{align}
For a projected planetary ellipse with $A = R_{\rm eq}$ and
$B = R_{\rm eq}(1-f_\perp)$, expansion to first
order in $f_\perp$ gives
\begin{align}
    L \approx 2R_{\rm eq}\left(1-\frac{f_\perp}{2}-\frac{f_\perp}{2}\cos2\phi\right).
    \label{eq:Lmax_approx}
\end{align}
This approximation is accurate for realistic giant
planets: even for $f_\perp=0.1$, comparable to Saturn's
flattening, the fractional error in $L$ is $\lesssim0.4\%$.

The light-curve asymmetry is controlled, to leading order,
by the fractional difference between the characteristic chord lengths at ingress and egress. Thus, the geometric factor $F$ in Equation~\ref{eq:Delta_BIC_Oblateness_1} obeys
\begin{align}
    F \propto \frac{L_{\rm ing}-L_{\rm egr}}{2R_{\rm eq}}.
    \label{eq:diffL}
\end{align}
Let $\theta_\perp$ be the projected obliquity, measured 
from the transit chord to the projected spin axis, and let
$\theta_\parallel$ be the angle between the
transit chord and the local normal to the stellar limb at the point
where the planet's center crosses over the limb.
For a circular stellar disk, $\theta_\parallel = \sin^{-1} b$.
The limb orientations at ingress and egress are
\begin{align}
    \phi_{\rm ing} &= \theta_{\perp}+\theta_{\parallel},\\
    \phi_{\rm egr} &= \theta_{\perp}-\theta_{\parallel}.
\end{align}
Substituting Equation~\ref{eq:Lmax_approx} into Equation~\ref{eq:diffL} 
gives
\begin{equation}
    F(f_\perp,\theta_{\perp}, b)
    \propto \frac{f_\perp}{2}\left[\cos2(\theta_{\perp}+\theta_{\parallel})-\cos2(\theta_{\perp}-\theta_{\parallel})\right] 
    \propto f_\perp\,\sin2\theta_{\perp}\sin2\theta_{\parallel}.
    \label{eq:geom_f}
\end{equation}
The sign of $F$ determines which of ingress or egress is steeper; the
detectability metric depends on $F^2$. Thus, 
to leading order, the asymmetry scales linearly with projected flattening.
It is also modulated by both $\theta_{\perp}$, the projected spin orientation,
and $\theta_{\parallel}$, which depends on transit geometry.

\section{Validation of Detectability Metrics}\label{app:detectability_scalings}

We calibrated the detectability metrics presented in Section~\ref{sec:metrics}
and Appendix~\ref{app:G-factor-derivation}
with injection-recovery simulations.
Synthetic transit light curves for oblate planets and planet-moon systems were generated with \texttt{squishyplanet}\footnote{Available at \href{https://github.com/ben-cassese/squishyplanet}{https://github.com/ben-cassese/squishyplanet}.} \citep{cassese2024} and \texttt{pandora}\footnote{Available at \href{https://github.com/hippke/pandora}{https://github.com/hippke/pandora}.} \citep{hippke2022}.
The light curves were sampled at 1-minute cadence and injected with Gaussian
noise using the same ETC-based and empirical JWST noise models adopted in the main text. Each simulated light curve was fitted with both the signal model and the corresponding null model using the Levenberg-Marquardt algorithm, freely varying quadratic limb darkening coefficients, $a$, $b$, transit time, baseline flux, and transit depth. The oblate model additionally fits for the flattening factor $f$ and obliquity $\theta$, while the planet$+$moon model additionally fits for the moon's radius, orbital phase, and orbital period. We placed the moon at the orbital phase farthest from the planet, so that the moon transit is separated from the planetary transit and the moon’s sky-projected velocity is approximately the planetary velocity. Detectability was quantified by the resulting $\Delta\chi^2$.

\setcounter{figure}{0}
\renewcommand{\thefigure}{B\arabic{figure}}
\begin{figure*}[t]
    \centering
    \includegraphics[width=1.0\linewidth]{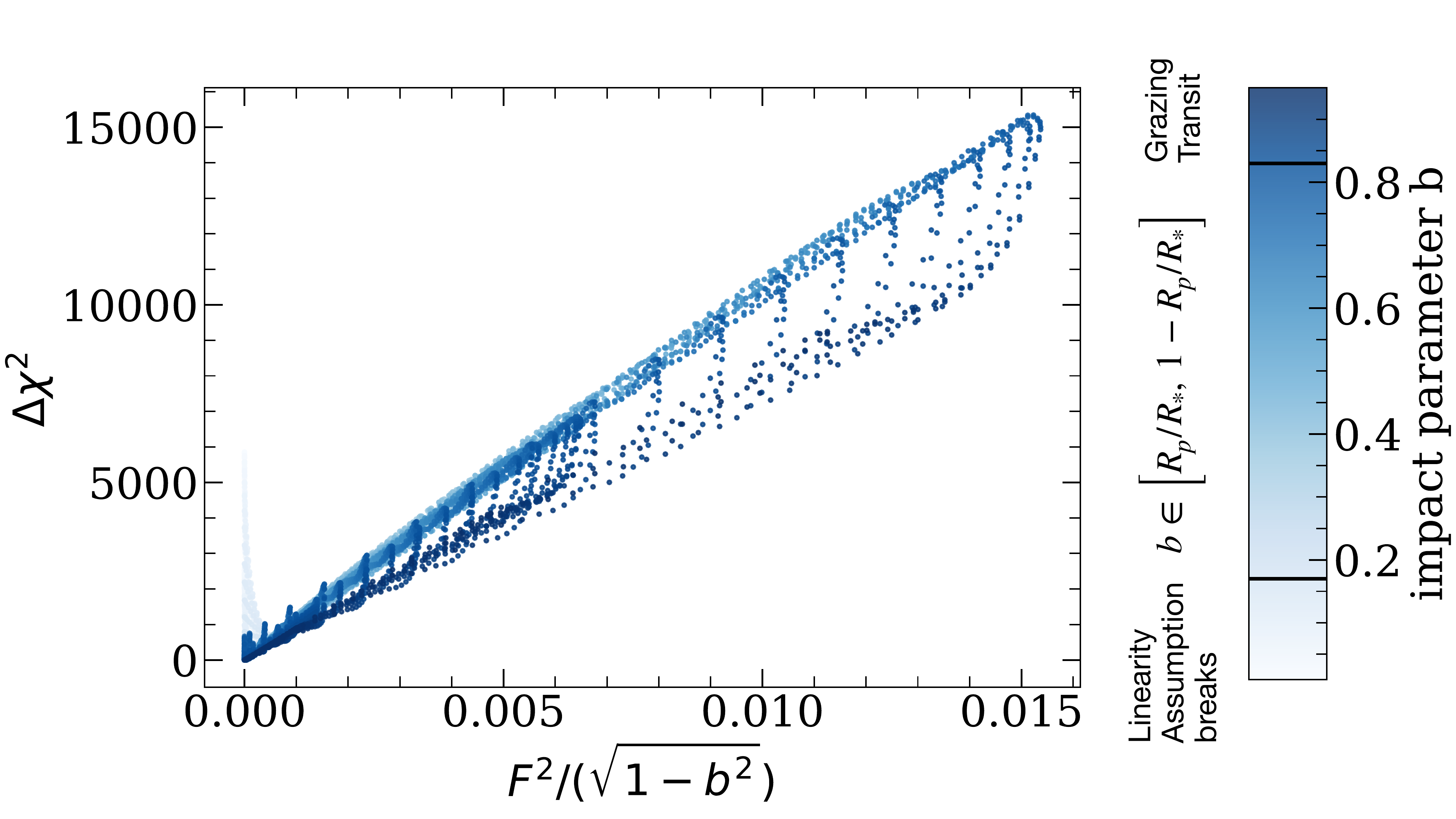}
    \caption{Validation of the analytic
    geometric scaling for oblateness detectability. Each point shows a simulated configuration for a Jupiter-like planet transiting a $0.60,R_\odot$ star, plotted as the oblateness-model preference $\Delta\chi^2$ versus the combined geometric predictor $F^2/\sqrt{1-b^2}$, where $F\equiv F(f_\perp,\theta_\perp,b)$ is the dimensionless ingress-egress asymmetry factor. Points are colored by impact parameter.
    For non-grazing and non-central
    transits, $\Delta\chi^2$ increases approximately linearly with the predicted scaling. The relation breaks down for central transits, where the
    ingress-egress asymmetry is suppresssed,
    and for grazing transits, where
    finite-planet-size effects reduce the
    asymmetry. For this system, these regimes occur at $b\lesssim 0.17$ and $b\gtrsim 0.83$, indicated by the markers on the color bar.}
    \label{fig:geom_f}
\end{figure*}

The simulations spanned a range of values for all parameters
of the metrics.
For oblateness, we sampled impact parameters $0<b<0.95$,
projected obliquities $0^\circ< \theta_\perp < 180^\circ$,
projected flattenings $f_\perp\in\{0.03,0.06,0.1\}$,
and planet radii $0.6 < R_p/R_J < 1.8$.
For moons, we sampled moon radii in units
of Ganymede's radius from 0.6--9.2, with additional
checks at $\{1,5,10\}$.
For both calculations, we sampled orbital distances
from 0.33 to 2~AU,
apparent magnitudes from $K=7$ to 14,
and stellar effective temperatures from 4000 to 
7200\,K.
To keep the stellar properties physically self-consistent, we treated the host stars as a one-parameter sequence along the zero-age main sequence, assigning $R_\star$ and $M_\star$ from the \citet{PecautMamajek2013} relations.

Figure~\ref{fig:geom_f} shows the key geometric validation for oblateness. For 
transits that are neither nearly central
nor nearly grazing,
the simulated $\Delta\chi^2$ is approximately linear in $F^2/\sqrt{1-b^2}$, as expected
from Equation~\ref{eq:oblateness_general_circular}.
The departures at small $b$ occur
because ingress and egress sample nearly the same
local limb orientation, weakening the asymmetry.
The departures at large $b$ occur because
the transit becomes grazing
and the local straight-limb
approximation breaks down.

The following expressions use the
dimensionless scalings of Section~\ref{sec:metrics}
with an overall normalization fitted to the
injection-recovery simulations.
The reference noise values of 78.2~ppm
and 76.8~ppm are the values of the ETC-based
and empirical noise models at the chosen
reference magnitude.
\begin{align}
\Delta \chi^2_{\rm ETC} \simeq
 108 &\left(\frac{f_\perp}{0.065}\right)^{2}\sin^{2}\!\bigl(2\theta_\perp\bigr)\nonumber\\
&\times\sin^{2}\!\left[
  2\arctan\!\left(\frac{b}{\sqrt{1-b^{2}}}\right)
\right]
\left(1-b^{2}\right)^{-1/2} \nonumber\\
&\times
\left(\frac{R_p}{R_{J}}\right)^{5}
\left(\frac{R_\star}{R_\odot}\right)^{-4}
\left(\frac{a}{1\,\mathrm{AU}}\right)^{1/2}
\left(\frac{M_\star}{M_\odot}\right)^{-1/2}\nonumber\\
&\times
\left(\frac{\sigma_{\rm ETC}(K)}{78.2}\right)^{-2},
\label{eq:oblateness_empirical_etc_app}
\end{align}
and
\begin{align}
\Delta \chi^2_{\rm emp} \simeq
 114 &\left(\frac{f_\perp}{0.065}\right)^{2}\sin^{2}\!\bigl(2\theta_\perp\bigr)\nonumber\\
&\times\sin^{2}\!\left[
  2\arctan\!\left(\frac{b}{\sqrt{1-b^{2}}}\right)
\right]
\left(1-b^{2}\right)^{-1/2} \nonumber\\
&\times
\left(\frac{R_p}{R_{J}}\right)^{5}
\left(\frac{R_\star}{R_\odot}\right)^{-4}
\left(\frac{a}{1\,\mathrm{AU}}\right)^{1/2}
\left(\frac{M_\star}{M_\odot}\right)^{-1/2}\nonumber\\
&\times
\left(\frac{\sigma_{\rm emp}(K)}{76.8}\right)^{-2}.
\label{eq:oblateness_empirical_real_app}
\end{align}
Across the full suite of simulations,
the calibrated prediction formula
agrees well with the
results of fitting simulated light curves.
A power-law fit gives
$\ln\Delta\chi^2_{\rm measured}\propto \ln\Delta\chi^2_{\rm pred}$ with slope $1.02$ for both noise models.
Thus the semi-analytic metric captures the
dominant dependence on geometry, planet radius, stellar properties,
orbital distance, and photometric precision, apart from the central and
grazing regimes identified above.

For moons, the calibrated predictors are
\begin{align}
    \Delta\chi^2_{\rm ETC} \;\approx\;
&24\,
\left(\frac{R_m}{R_{\text{Ganymede}}}\right)^{4}\nonumber\\
&\times
\left(\frac{R_\star}{R_\odot}\right)^{-3}
\left(\frac{M_\star}{M_\odot}\right)^{-1/2}\sqrt{1-b^2}\nonumber\\
&\times
\left(\frac{a}{1~\mathrm{AU}}\right)^{1/2}
\left(\frac{\sigma_{\rm ETC}(K)}{78.2}\right)^{-2},
\label{eq:moon_empirical_etc_app}
\end{align}
and
\begin{align}
    \Delta\chi^2_{\rm emp} \;\approx\;
&25\,
\left(\frac{R_m}{R_{\text{Ganymede}}}\right)^{4}\nonumber\\
&\times
\left(\frac{R_\star}{R_\odot}\right)^{-3}
\left(\frac{M_\star}{M_\odot}\right)^{-1/2}\sqrt{1-b^2}\nonumber\\
&\times
\left(\frac{a}{1~\mathrm{AU}}\right)^{1/2}
\left(\frac{\sigma_{\rm emp}(K)}{76.8}\right)^{-2}.
\label{eq:moon_empirical_real_app}
\end{align}
Equivalently, the ETC-based moon metric may be written in duration form as
\begin{align}
\Delta\chi^2_{\rm ETC}\approx
1.87\left(\frac{R_m}{R_{\rm Ganymede}}\right)^4
\left(\frac{R_\star}{R_\odot}\right)^{-4}
\left(\frac{T}{1\,{\rm hr}}\right)
\left(\frac{\sigma_{\rm ETC}(K)}{78.2}\right)^{-2} \label{eq:moon_empirical_etc_used},
\end{align}
with an analogous expression for the empirical noise model. The results of the analytic estimate
and light-curve fitting
follow the expected scaling with $R_m^4$, transit duration, stellar radius, and noise level,
with no significant residual trends over the parameter ranges tested.

These calibrated expressions are the detectability metrics used throughout the forecast calculations in the main text.

\bibliography{refs}{}
\bibliographystyle{aasjournalv7}

\end{CJK*}
\end{document}